\documentclass[aps,twocolumn]{revtex4}
\makeatletter

\newcommand{\Rmnum}[1]{\expandafter\@slowromancap\romannumeral #1@}
\makeatother
\usepackage{makecell}
\usepackage{graphicx}
\usepackage{dcolumn}
\usepackage{bm}
\usepackage{CJK}
\usepackage{amsfonts}
\usepackage{psfrag}
\usepackage{wrapfig}
\usepackage{subfigure}
\usepackage{makeidx}
\usepackage{multirow}
\usepackage{epsf}
\usepackage{amsmath}
\usepackage{cases}
\usepackage{multirow}
\usepackage{amssymb}
\usepackage{soul}
\usepackage[version=3]{mhchem}

\usepackage[colorlinks,linkcolor=red,anchorcolor=red,citecolor=blue,urlcolor=blue]{hyperref}

\begin{document}
\title{Beating solitons with nonzero relative wavenumber, their interaction, and coexistence with Akhmediev breathers}
\author{Wen-Juan Che$^{1}$}
\author{Chong Liu$^{1,2,3}$}\email{chongliu@nwu.edu.cn}
	
\address{$^1$School of Physics, Northwest University, Xi'an 710127, China}
\address{$^2$Shaanxi Key Laboratory for Theoretical Physics Frontiers, Xi'an 710127, China}
\address{$^3$Peng Huanwu Center for Fundamental Theory, Xi'an 710127, China}
\begin{abstract}
We study beating solitons and their interaction for the three-component nonlinear Schr\"odinger equations (NLSEs) in both focusing and defocusing cases.
We identify a previously unknown family of beating solitons with nonzero relative wavenumber which is absent in the two-component NLSEs.
Existence diagrams of these beating solitons are extracted from the exact solutions.
Unique coexistence and interaction between beating solitons and vector Akhmediev breathers are revealed by using the existence diagrams. It is demonstrated that the coexisting breathers turn out to be the two-component Akhmediev breathers with nonzero relative wavenumber.
The validity of our exact solutions has been confirmed by numerical simulations.

\end{abstract}

\maketitle
\section{INTRODUCTION}\label{Sec-Intro}

Nonlinear interaction of multicomponent wave fields is a common phenomenon in physics \cite{OF,Chen,BEC,PT,F1,F2}. One of the most paradigmatic models that describes such interaction within integrable nonlinear systems is presently known as Manakov equations \cite{MM}. Despite a seemingly simple vector generalization of the nonlinear Schr\"odinger equation (NLSE), Manakov equations play a pivotal
role in describing more complex wave phenomena in hydromechanics \cite{F1}, in optics \cite{OF,Chen}, and in Bose-Einstein condensates \cite{BEC}.
One of them is oscillating wave structures on a plane wave background \cite{Book97}. Also known as `breathers', they can take large variety of different forms. Among well known examples of such breathers are beating solitons \cite{Park-2000} which are absent in the scalar NLSE.
The distinctive feature of beating solitons is that they exhibit oscillating behavior in individual components; while their total intensity of all components shows plain (nonoscillating) soliton profile.
Such property has been confirmed experimentally in two-component Bose-Einstein condensates \cite{Hoefer-2011}.

One way to obtain beating solitons is to use the SU(2) rotation symmetry for dark-bright soliton solutions of the two-component NLSEs \cite{Park-2000}. These beating solitons strongly depends on the choice of the dark-bright solitons. Thus, only special types are obtained.
Recent numerical simulations confirmed the robustness of these beating solitons even under the non-integrable condition \cite{Wang-2016,Wang-2021}. Understanding such excitations from the point of view of quantum mechanics has also been proposed  \cite{Zhao-2018}.

An alternative approach is to construct beating soliton solutions from the vector plane wave via either the Ba\"{c}klund transformation \cite{OCW1,OCW2,OCW3} or the Darboux transformation (or dressing method) \cite{VSRB-2023,VBE-2024}. Using such approach and combining the eigenvalue analysis, a full classification of beating solitons in both the focusing and defocusing regimes is given \cite{VBE-2024}.
Based on these solutions, nontrivial spectral asymmetry of moving beating solitons is revealed.
Another advantage of these beating solitons lies in the variety of  possibilities of their nonlinear superposition.
It is shown that collisions of certain types of beating solitons can induce the scalar NLSE breather patterns in the focusing regime \cite{VSRB-2023}.
Among them, we can mention the beating-soliton-collision-induced Akhmediev breathers (ABs) which provide a new way to generate the modulation instability \cite{VBE-2024}. Moreover, these results can be used to generate beating stripe solitons arising from helicoidal spin-orbit coupling in Bose-Einstein condensates \cite{Qin-2025}.

It should be pointed out that in all these findings mentioned above, the zero relative wavenumber of the plane wave is a prerequisite for the existence of beating solitons \cite{VSRB-2023,VBE-2024}.
Indeed, recent works \cite{VKMS-f2022, VKMS-df2023} demonstrated that beating solitons can be a special case of vector Kuznetsov-Ma solitons (KMSs) with zero relative wavenumber.

In this paper, we present a new family of beating solitons with nonzero relative wavenumber by extending the study of beating solitons to the three-component NLSEs.
Although localized waves of three-component NLSEs including vector solitons \cite{G-2016,Kanna_NLSE3,SC0,SC1,SC2,Qin_NLSE3}, vector rogue waves \cite{VB5}, and vector breathers \cite{VB8} have been studied extensively, these beating solitons have not been reported before.
We show that such excitations exhibit a unique beating-dark-beating soliton structure in both focusing and defocusing cases which are absent in the two-component NLSEs \cite{Park-2000,Zhao-2018,VSRB-2023,VBE-2024}. We establish a link by showing that these beating solitons can also be obtained by using SU(2) symmetry for the dark-dark-bright solitons.
Existence diagrams of these beating solitons are extracted from the exact solutions. Using the existence diagrams, we reveal the previously unknown coexistence and interaction between beating solitons and vector ABs. It is demonstrated that the coexisting breathers turn out to be the ABs of two-component NLSEs with nonzero relative wavenumber. As these ABs exhibit highly nontrivial dynamics in nonlinear stage of modulation instability \cite{VAB-2021,VAB-Df2022,VAB-2022,VAB-2023}, our results could provide a way to generate complex vector modulation instability pattern by using beating soliton collisions.
On the other hand, the physical relevance of three-component NLSEs has been confirmed recently in experiment \cite{three-2020} by
observing the bright-dark-bright solitons in Bose-Einstein condensates with repulsive forces between the atomic components. Our present theoretical results may provide a basis for observing more complex wave patterns in such experiments.

The paper is organized as follows. In Section \ref{Sec-Model}, we present exact fundamental solutions of the three-component NLSEs which contain as a special case, the beating soliton solution with nonzero relative wavenumber.
We show that in this case, the model can reduce to the two-component NLSEs that admit nontrivial vector breather solutions in Section \ref{Sec-AB}.
Section \ref{Sec-b2} presents the characteristics of beating solitons with nonzero relative wavenumber and coexisting ABs for both defocusing and focusing cases. In Section \ref{Sec-su}, we establish the link by showing that these beating solitons can also be obtained by using SU(2) symmetry for the dark-dark-bright solitons.
Section \ref{Sec-Second} explores second-order nonlinear superpositions and the coexistence of two fundamental beating solitons with vector ABs. Numerical simulations and conclusions are presented in Sections \ref{Sec-Numerical} and \ref{Sec-Conc}, respectively.

\section{Model and fundamental solution}\label{Sec-Model}

We consider the three-component NLSEs in both focusing and defocusing regimes in the following dimensionless form:
\begin{equation}\label{eq1}
i\frac{\partial\psi^{(j)}}{\partial t}+\frac{1}{2}\frac{\partial^2\psi^{(j)}}{\partial x^2}+\delta(\sum^3_{n=1}|\psi^{(n)}|^2)\psi^{(j)}=0,
\end{equation}
where $\psi^{(j)}(t,x)$ $(j=1,2,3)$ are the three nonlinearly coupled components of the vector wave field. The physical meaning of independent variables $x$ and $t$ depends on a particular physical problem of interest.
We have normalized Eqs. (\ref{eq1}) in a way such that $\delta=\pm1$.
Note that in the case $\delta=1$, Eqs. (\ref{eq1}) refer to the focusing regime; in the case $\delta=-1$, Eqs. (\ref{eq1}) refer to the defocusing regime.

In order to obtain the beating soliton solution, we shall start with the plane wave solution $\psi_{0}^{(j)}$ as the seed solution, which is given by:
\begin{equation}
\psi_{0}^{(j)}=a_j\exp{\{i\theta_j\}},\label{eqpw0}
\end{equation}
where
\begin{equation}
\theta_j=\beta_j x + \delta(a_1^2+a_2^2+a_3^2- \frac{1}{2}\beta_j^2 )t.\label{eqpw}
\end{equation}
Here, $a_j$ and $\beta_j$ are the amplitudes and wavenumbers of the $\psi^{(j)}$ component, respectively.
The wavenumbers $\beta_j$ play a key role in beating soliton formation.
A nontrivial generalization we show here is that beating solitons can exist on the vector plane wave with nonzero relative wavenumber. Namely, we consider the plane wave (\ref{eqpw0}) with the wavenumbers
\begin{equation}
\beta_1=\beta_3=\beta\neq0,~\beta_2=0. \label{eqwnb}
\end{equation}
Moreover, to simplify our analysis we take here
\begin{equation}
a_1=a_3=a/\sqrt{2},~a_2=a. \label{eqpwa}
\end{equation}
This condition is also convenient for us to analyze the degeneration of the model (\ref{eq1}), see details in Section \ref{Sec-AB}.
Using the plane wave (\ref{eqpw0}) under the conditions (\ref{eqwnb}) and (\ref{eqpwa}) as the seed solution, we obtain the vector fundamental (first-order $n=1$) solution by performing the Darboux transformation shown in \cite{VBE-2024}. Namely, the solutions describing waves on the background (\ref{eqpw0}) with the conditions (\ref{eqwnb}) and (\ref{eqpwa}) are given by
\begin{eqnarray}\label{eqdt1}
\begin{split}	&\psi_{}^{(1)}=\psi_{0}^{(1)}+\frac{(\lambda_{1}^*({\chi}_{a})-\lambda_{1}({\chi}_{a}))R_{1}^*S_{1}}{|R_{1}|^2+\delta|S_{1}|^2+\delta|W_{1}|^2+\delta|X_{1}|^2},\\	&\psi_{}^{(2)}=\psi_{0}^{(2)}+\frac{(\lambda_{1}^*({\chi}_{a})-\lambda_{1}({\chi}_{a}))R_{1}^*W_{1}}{|R_{1}|^2+\delta|S_{1}|^2+\delta|W_{1}|^2+\delta|X_{1}|^2}.\\
&\psi_{}^{(3)}=\psi_{0}^{(3)}+\frac{(\lambda_{1}^*({\chi}_{a})-\lambda_{1}({\chi}_{a}))R_{1}^*X_{1}}{|R_{1}|^2+\delta|S_{1}|^2+\delta|W_{1}|^2+\delta|X_{1}|^2}.\\
\end{split}
\end{eqnarray}
Details of the calculation are given in Appendix \ref{Sec-0}. Here, $\lambda_{1}$ denotes the spectral parameter of the Lax pair, Eq. (\ref{eqlambda}), and ${\chi}_{a}$ is the corresponding eigenvalue, Eq. (\ref{b-eqchi1234}). $R_{1}$, $S_1$, $W_{1}$, $X_1$ are the eigenfunctions consisting of trigonometric and hyperbolic functions, which are given by Eq. (\ref{EqH3-2}).

On the other hand, just as the previous results \cite{VKMS-f2022,VKMS-df2023,VBE-2024}, one can also construct the solutions describing beating solitons with the same wavenumber $\beta_1=\beta_2=\beta_3$. We remark that the resulting beating solitons are the trivial generalization of two-component NLSEs \cite{VBE-2024,VSRB-2023}. We show the details of this case in Appendix \ref{Sec-4}.

Solutions (\ref{eqdt1}) depend on the parameters of plane wave background ($a$, $\beta$), the real constants ($\alpha$, $\gamma$), the nonlinearity coefficient $\delta$, and the coefficients of the vector eigenfunction $\{c_{1,a}, c_{1,b}, c_{1,c}, c_{1,d}\}$.
Once $a$ and $\beta$ are fixed, the parameters ($\alpha$, $\gamma$), $\{c_{1,a}, c_{1,b}, c_{1,c}, c_{1,d}\}$, and the choice of ${\chi}_{a}$  play a key role in the fundamental breather (or soliton) formation.

General speaking, single breathers or beating solitons exist when two of $\{c_{1,a}, c_{1,b}, c_{1,c}, c_{1,d}\}$ vanishes.
Single breathers exist when $\{c_{1,a}, c_{1,b}, 0, 0\}$ and beating solitons for the rest.
Classification for the beating solitons is given in Table .
When three of these real constants in $\{c_{1,a}, c_{1,b}, c_{1,c}, c_{1,d}\}$ are zero, Eqs. (\ref{eqdt1}) reduce the plane wave solution. When only one of these real constants $\{c_{1,a}, c_{1,b}, c_{1,c}, c_{1,d}\}$ are zero or neither are zero, Eqs. (\ref{eqdt1}) produces a coexistence of multiple localized waves.
Changing the values of nonzero real parameters $\{c_{1,a}, c_{1,b}, c_{1,c}, c_{1,d}\}$ affects only the positions of beating solitons in the ($x-t$) plane.

\section{Degeneration of the model and breather solutions}\label{Sec-AB}

In addition to the classification of solutions (\ref{eqdt1}) shown above,
we remark that the solutions of the single breathers should be analyzed more carefully. As we will show below that the single breather solutions under the conditions (\ref{eqwnb}) and (\ref{eqpwa}) can be reduced to the solutions of two-component NLSEs. We call this process `degeneration' of the model (\ref{eq1}). The details are as follows.

Let us consider the single breather solutions (\ref{eqdt1}) with $\{c_{1,a}, c_{1,b}, c_{1,c}, c_{1,d}\}=\{1, 1, 0, 0\}$. Then, we have $S_1=X_1$ from (\ref{eqdt1}), which means that
\begin{equation}\label{eqsb1}
\psi^{(1)}=\psi^{(3)}.
\end{equation}
If we consider the scaling
\begin{equation}\label{eqsb2}
\psi^{(1)}=\tilde{\psi}^{(1)}/\sqrt{2},
\end{equation}
the three-component NLSEs (\ref{eq1}) then reduce to
\begin{equation}\label{eqsb3}
	\begin{split}
		i\frac{\partial\tilde{\psi}^{(1)}}{\partial t}+\frac{1}{2}\frac{\partial^2\tilde{\psi}^{(1)}}{\partial x^2}+\delta(|\tilde{\psi}^{(1)}|^2+|\psi^{(2)}|^2)\tilde{\psi}^{(1)}&=0,\\
		i\frac{\partial\psi^{(2)}}{\partial t}+\frac{1}{2}\frac{\partial^2\psi^{(2)}}{\partial
x^2}+\delta(|\tilde{\psi}^{(1)}|^2+|\psi^{(2)}|^2)\psi^{(2)}&=0.
	\end{split}
\end{equation}
This is the standard two-component NLSEs (Manakov equations) where $\tilde{\psi}^{(1)}$ and $\psi^{(2)}$ describe the vector breathers on the vector plane wave $(\tilde{\psi}_0^{(1)}, \psi_0^{(2)})$. The latter are given by
\begin{equation}\label{eqsb4}
\begin{split}
&\tilde{\psi}_0^{(1)}=\sqrt{2}\psi_0^{(1)}=a \exp{\{i[\beta x + (2\delta a^2-\frac{1}{2}\beta^2 )t]\}}.\\
&\psi_0^{(2)}=a\exp{\{i[2\delta a^2t]\}},
\end{split}
\end{equation}
respectively.
As can be seen from (\ref{eqsb4}), the new plane waves $\tilde{\psi}_0^{(1)}$ and $\psi_0^{(2)}$ have the equal amplitude $a$ and the relative wavenumber $\beta$.

Equations (\ref{eqsb1})-(\ref{eqsb4}) show clearly how the model (\ref{eq1}) reduces to the two-component NLSEs under the condition of single breather solutions. This result will help us understand the interaction of beating solitons with nonzero relative wavenumber in Section \ref{Sec-Second}.

On the other hand, as shown in Refs. \cite{VB4,VB7,VRW-2016,VAB-2021,VAB-Df2022,VRW-2024,VAB-2022,VRW-2022,VAB-2023,VSRB-2024}, the vector breathers described by $(\tilde{\psi}^{(1)}, \psi^{(2)})$ are nontrivial vector generalization due to the nonzero relative wavenumber $\beta\neq0$. We then expect the unique coexistence and interaction with the beating solitons.
This is the main finding when we consider the beating solitons with nonzero wavenumber which has never been reported before.

Moreover, the vector breathers ($\tilde{\psi}^{(1)}$, $\psi^{(2)}$) contain the vector general breathers ($\alpha\neq0$, $\gamma\neq0$), the vector ABs ($\alpha=0$, $\gamma\neq0$), and the vector KMSs
($\alpha\neq0$, $\gamma=0$). The expressions are given in Appendix \ref{Sec-GB}.
In the following, we focus our attentions on the beating solitons and their coexistence with ABs.

\section{Beating solitons with nonzero relative wavenumber and coexisting ABs}\label{Sec-b2}
Let us explore the case of beating solitons under the condition with the nonzero relative wavenumber $\beta_1=\beta_3=\beta$, $\beta_2=0$.
As shown in Table \ref{Table3} (see Appendix \ref{Sec-0}), we classify beating solitons by eigenvalues (${\chi}_{a,1}$, ${\chi}_{a,2}$, ${\chi}_{a,3}$, ${\chi}_{a,4}$) and $\{c_{1,a}, c_{1,b}, c_{1,c}, c_{1,d}\}$.
Moreover, there is a parameter $\chi_{k}$, which is related to the eigenvalue $\chi_{a}$, $\chi_{b}$ or $\chi_{c}$ depending on the special cases we consider. Thus, in the following, we denote the beating soliton solutions as  $\psi_{}^{(j)}(\chi_{a},\chi_k)$.

Specifically, fundamental beating solitons exist either when $\{c_{1,a}, 0, 0, c_{1,d}\}$, $\{0, c_{1,b}, 0, c_{1,d}\}$, $\{0, 0, c_{1,c}, c_{1,d}\}$ for certain ($\alpha$, $\gamma$). All types of solutions can be presented in a unified form:
\begin{eqnarray}\label{beating-s4}
\begin{split}
&\psi_{}^{(1)}(\chi_{a},\chi_k)=\sqrt{2}/2\left(\psi_{1}^{(1)}+\psi_{1}^{(3)}\right),\\
&\psi_{}^{(2)}(\chi_{a},\chi_k)=\psi_{1}^{(2)},\\
&\psi_{}^{(3)}(\chi_{a},\chi_k)=\sqrt{2}/2\left(\psi_{1}^{(1)}-\psi_{1}^{(3)}\right),\\
\end{split}
\end{eqnarray}
where
\begin{eqnarray}\label{DB-s4}
\begin{split}
&\psi_{1}^{(1)}=\psi_{0}^{(1)}\left(1+\frac{\mathcal{M}_k}{\epsilon_k}+\frac{\mathcal{M}_k}{\epsilon_k}\tanh(\Delta_k+\frac{1}{2}\ln{[\epsilon_k]})\right),\\
&\psi_{1}^{(2)}=\psi_{0}^{(2)}\left(1+\frac{\mathcal{L}_k}{\epsilon_k}+\frac{\mathcal{L}_k}{\epsilon_k}\tanh(\Delta_k+\frac{1}{2}\ln{[\epsilon_k]})\right),\\
&\psi_{1}^{(3)}=\exp{(i\theta_3)}\left(\frac{\mathcal{P}_k}{\sqrt{\epsilon_k}}
\textmd{sech}(\Delta_k+\frac{1}{2}\ln{[\epsilon_k]})\exp(-i\Omega_k)\right).\\
\end{split}
\end{eqnarray}
and
\begin{eqnarray}
\begin{split}
&\epsilon_k=\frac{1}{\delta}+\frac{a_1^2}{(\beta_1+\chi_k)(\beta_1+\chi_k^*)}+
\frac{a_2^2}{(\beta_2+\chi_k)(\beta_2+\chi_k^*)},\\
&\Delta_k=-\chi_{ki}(x+\chi_{kr}t),~~
\Omega_k=\chi_{kr}x+\frac{1}{2}(\chi_{kr}^2-\chi_{ki}^2)t.
\end{split}
\end{eqnarray}
The coefficients $\mathcal{M}_k$, $\mathcal{L}_k$ and $\mathcal{P}_k$ are respectively:
\begin{small}
\begin{eqnarray}
\begin{split}
\mathcal{M}_k=\frac{(\lambda^*-\lambda)}{2\delta(\beta_1+\chi_k)},~
\mathcal{L}_k=\frac{(\lambda^*-\lambda)}{2\delta(\beta_2+\chi_k)},
\mathcal{P}_k=\frac{(\lambda^*-\lambda)}{2\delta}.
\end{split}
\end{eqnarray}
\end{small}
Subscripts $r$ and $i$ denote the real and imaginary parts, respectively.
Note that, parameter $\chi_{k}$ is related to the eigenvalue $\chi_{a}$, $\chi_{b}$ or $\chi_{c}$ depending on the special cases we consider. Such a parameter plays a key role in the type of solitons. It also plays a key role in soliton dynamics. Namely, from solutions (\ref{beating-s4}),
we have the propagation velocity of fundamental solitons
\begin{equation}\label{eq-V}
Vg=-\chi_{kr}.
\end{equation}
The width of beating solitons in $x$ is
\begin{equation}\label{eq-W}
W=\frac{1}{|\chi_{ki}|}.
\end{equation}
Finally, the maximal soliton amplitudes are given by
\begin{eqnarray}\label{DB-}
\begin{split}
&|\psi_{k}^{(1)}|_{\textmd{max}}=|\psi_{k}^{(3)}|_{\textmd{max}}=\left|\frac{\sqrt{2}}{2}(1+\frac{\mathcal{M}_k}{\epsilon_k}+\frac{\mathcal{P}_k}{\sqrt{\epsilon_k}})\right|,\\
&|\psi_{k}^{(2)}|_{\textmd{max}}=\left|1+\frac{\mathcal{L}_k}{\epsilon_k}\right|.
\end{split}
\end{eqnarray}
Below, we illustrate the dynamics of beating solitons with nonzero relative wavenumber for the defocusing and focusing cases.
In particular, we consider the beating solitons that can coexist with vector ABs on the same plane wave.

\begin{figure}[htbp]
\centering
\includegraphics[width=84mm]{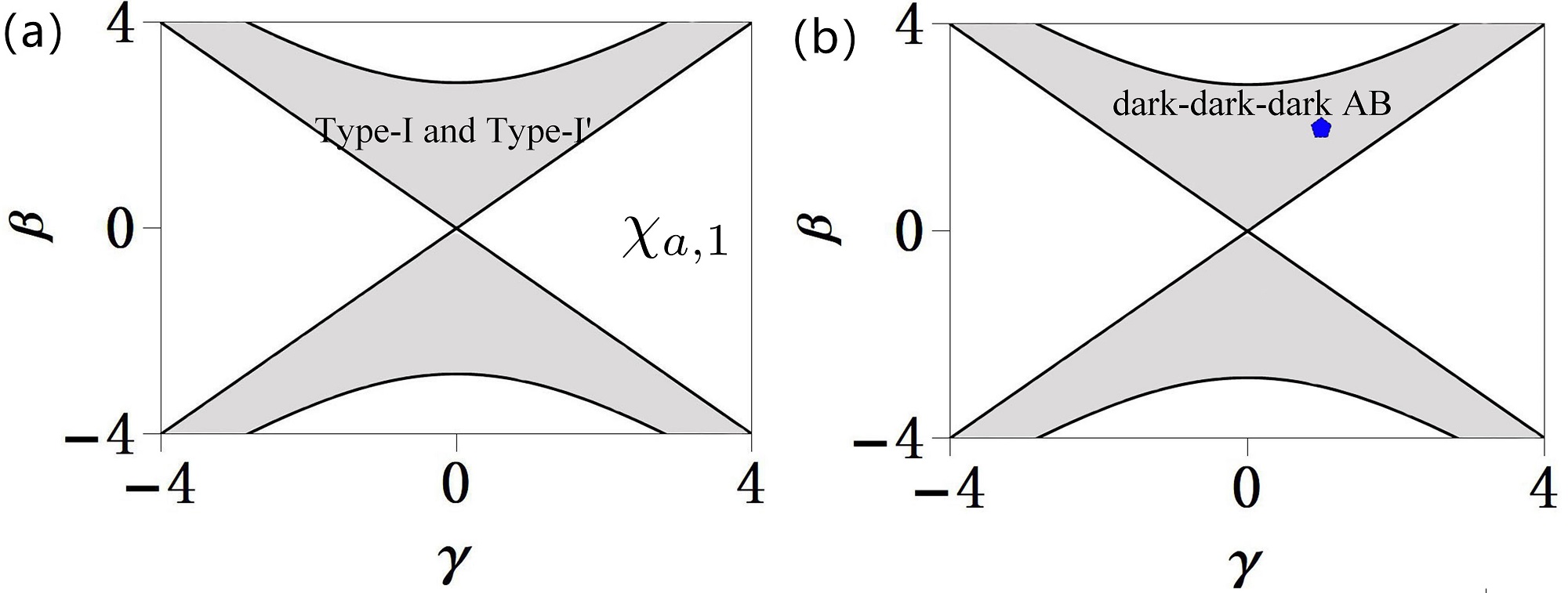}
\caption{Existence diagrams of type-$I$ and type-$I'$ beating solitons (a) and vector dark-dark-dark Akhmediev breathers (b) in the defocusing regime in the $(\gamma, \beta)$ plane with the eigenvalue $\chi_{a,1}$. The boundaries are given explicitly by Eq. (\ref{eq-edd}). The blue point corresponds to the solutions shown in Fig. \ref{f-d-beating2}.}
\label{f-d-phase-1}
\end{figure}

\begin{figure*}[htbp]
\centering
\includegraphics[width=150mm]{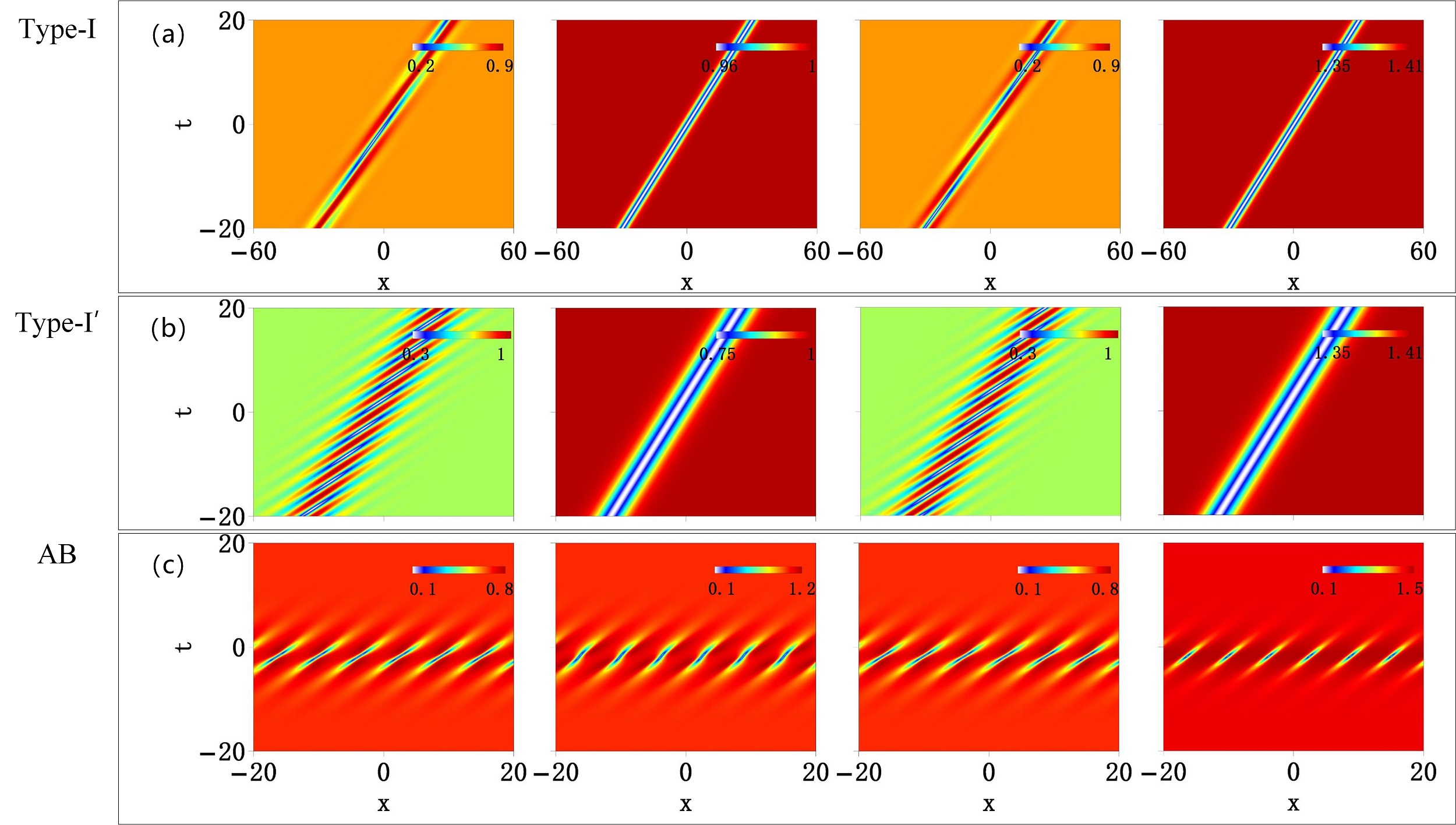}
\caption{Amplitude distributions of type-$I$ beating soliton (\ref{beating-s4}) (a) and type-$I'$ beating solitons (\ref{beating-s4}) (b), and vector dark Akhmediev breathers (c) in the defocusing regime. All these solutions correspond to the eigenvalue $\chi_{a,1}$. From left to right, $|\psi_{}^{(1)}|$, $|\psi_{}^{(2)}|$, $|\psi_{}^{(3)}|$, and the total amplitude $\sqrt{|\psi_{}^{(1)}|^2+|\psi_{}^{(2)}|^2+|\psi_{}^{(3)}|^2}$. Parameters are $a=1$, $\beta=2$, $\alpha=0$, and $\gamma=1$.}
\label{f-d-beating2}
\end{figure*}

\subsection{The defocusing regime}
First, we consider the solutions (\ref{beating-s4}) in the defocusing case $\delta=-1$.
Clearly, the choice of the parameters ($\alpha$,$\gamma$) and $\beta$ is not arbitrary. We need to perform the Hessian matrix analysis for (\ref{beating-s4}) in order to find the existence regions of these solutions. Using the technique presented in Ref. \cite{VKMS-f2022, VKMS-df2023}, we constructed the existence diagrams for all solutions given by (\ref{beating-s4}) corresponding to each of the eigenvalues.
Below we consider the case $\alpha=0$, $\gamma\neq0$.

Figure \ref{f-d-phase-1} illustrates the existence regions of solutions on the $(\gamma, \beta)$ plane.
Only two eigenvalues $\chi_{a,1}$, $\chi_{a,2}$ are valid in the defocusing regime, since $\chi_{a,3}$, $\chi_{a,4}$ are purely real.
The role of $\chi_{a,1}$ and $\chi_{a,2}$ in beating soliton (or breather) formation is the same.
For either $\chi_{a,1}$ or $\chi_{a,2}$, there two types of beating solitons existing in the X-shape (gray) area.
As an example, Fig. \ref{f-d-phase-1} (a) shows the diagram of (type-$I$ and type-$I'$) beating solitons for $\chi_{a,1}$. Namely, $\psi_1^{(j)}(\chi_{a,1},\chi_{a,1})$, $\psi_1^{(j)}(\chi_{a,1},\chi_{b,1})$.
In particular, the region also admits one vector AB in the defocusing regime. Importantly, we have the explicit expressions of the critical condition of the coexistence of beating solitons and ABs. They are given by
\begin{equation}\label{eq-edd}
\beta_{c1}=\pm\gamma,~~
\beta_{c2}=\pm\sqrt{8a^2+\gamma^2}.
\end{equation}
Thus, coexistence occurs in the region
\begin{equation}\label{}
\beta^2_{c1}<\beta^2<\beta^2_{c2}.
\end{equation}
This condition coincides with that of vector ABs of two-component NLSEs in the defocusing regime \cite{VAB-Df2022}.

Figure \ref{f-d-beating2} shows the amplitude profiles of two types of  beating solitons and the coexisting ABs. As can be seen, the three-component beating solitons exhibit oscillating properties in the $\psi^{(1)}$ and $\psi^{(3)}$ components, while the $\psi^{(2)}$ component is a non-oscillating dark soliton. As the amplitude distribution of $\psi^{(1)}$ and $\psi^{(3)}$ components is complementary, the total intensity of the wave field is still a non-oscillating dark soliton. This comes from the nature of defocusing regime.
Moreover, the two types of beating solitons differ in their oscillating period, velocity, and amplitudes.
However, their total intensity has the same amplitude distribution. The coexisting AB is shown in Fig. \ref{f-d-beating2}(c).
It shows a vector dark-dark AB.
Coexistence between two types of beating solitons and the breathers is shown in Section \ref{Sec-Second} below.

It should be pointed out that the coexistence between beating solitons and vector ABs reported above cannot occur in the case of the two-component defocusing NLSEs \cite{VBE-2024} and the case of three-component defocusing NLSEs with the zero relative wavenumber.

\begin{figure}[htbp]
\centering
\includegraphics[width=84mm]{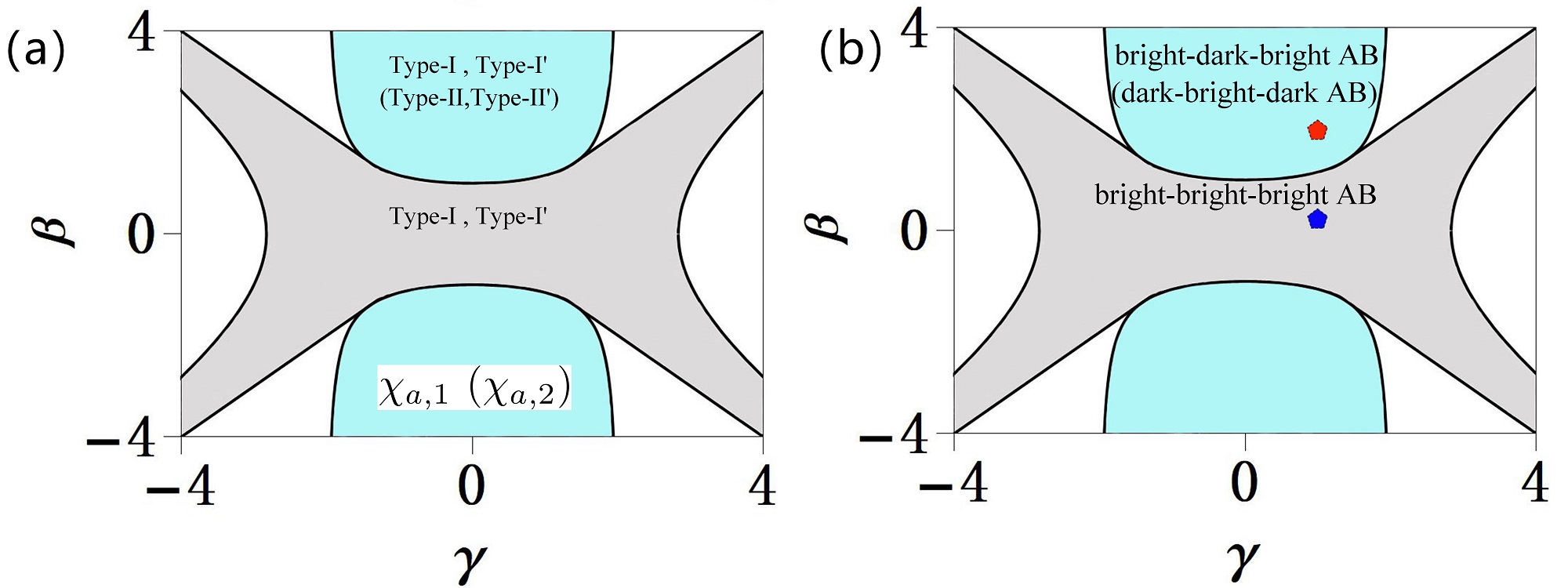}
\caption{Existence diagrams of beating solitons (a) and vector Akhmediev breathers (b) in the focusing regime in the $(\gamma, \beta)$ plane. The boundaries are given explicitly by Eqs. (\ref{eq-efd1}) and (\ref{eq-efd2}). In the cyan region, type-$I$ and type-$I'$ beating solitons and bright-dark-bright Akhmediev breather correspond to the eigenvalues $\chi_{a,1}$; while type-$II$ and type-$II'$ beating solitons and dark-bright-dark Akhmediev breather correspond to the eigenvalues $\chi_{a,2}$. In the gray region, the role of $\chi_{a,1}$ and $\chi_{a,2}$ in wave formation is the same, only type-$I$ and type-$I'$ beating solitons and bright-bright-bright Akhmediev breather exist. The marked blue (red) point corresponds to the solutions shown in Fig. \ref{f-f-beating2-a} (Figs. \ref{f-f-beating2-b} and \ref{f-f-beating2-c}).}
\label{f-f-phase-1}
\end{figure}

\begin{figure*}[htbp]
\centering
\includegraphics[width=150mm]{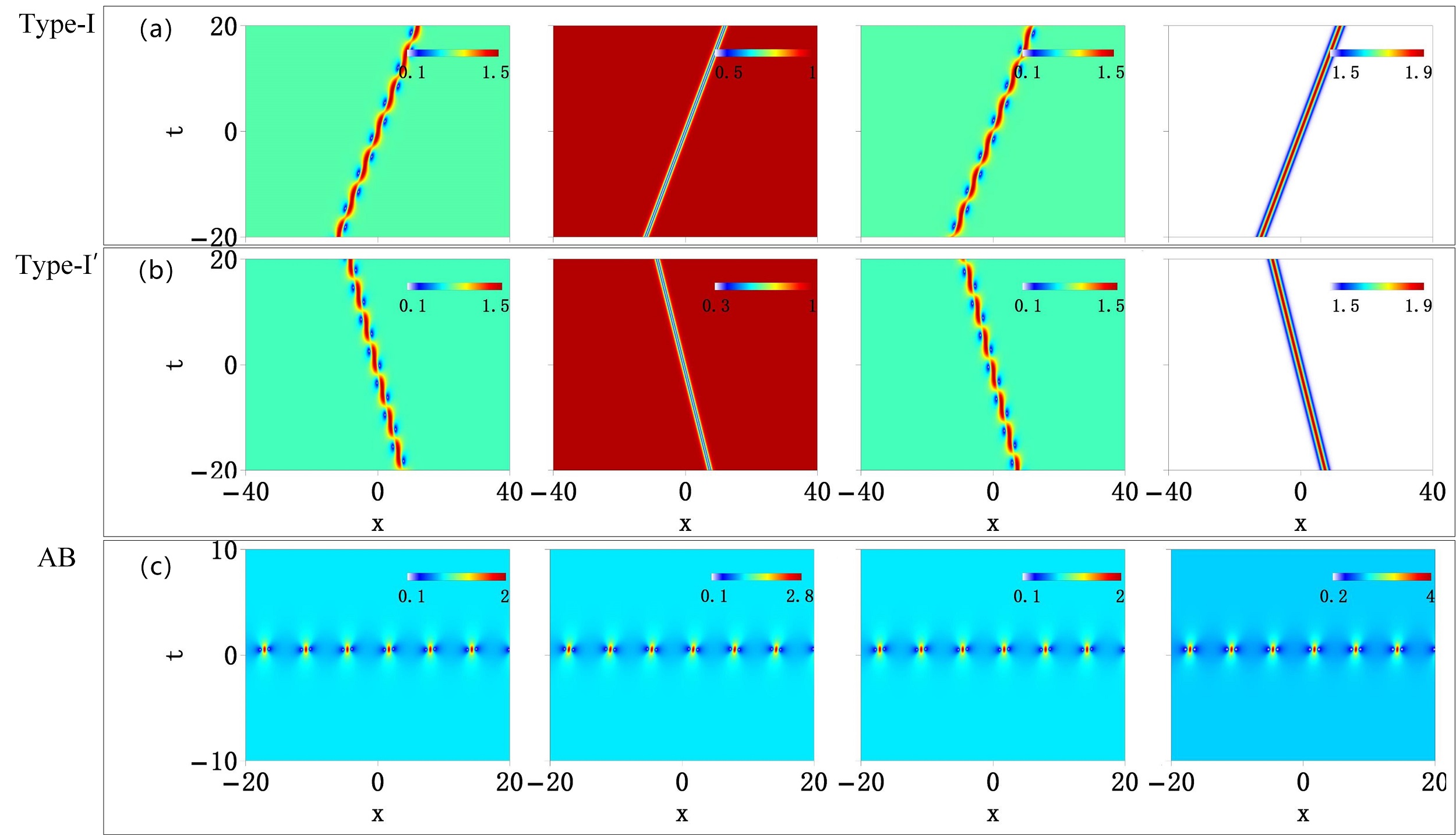}
\caption{Amplitude distributions of type-$I$ beating soliton (\ref{beating-s4}) (a) and type-$I'$ beating solitons (\ref{beating-s4}) (b), and vector bright-bright-bright Akhmediev breathers (c) in the focusing regime. All these solutions correspond to the eigenvalue $\chi_{a,1}$.  From left to right, $|\psi_{}^{(1)}|$, $|\psi_{}^{(2)}|$, $|\psi_{}^{(3)}|$, and the total amplitude $\sqrt{|\psi_{}^{(1)}|^2+|\psi_{}^{(2)}|^2+|\psi_{}^{(3)}|^2}$. Parameters are $a=1$, $\beta=0.2$, $\alpha=0$ and $\gamma=1$.}
\label{f-f-beating2-a}
\end{figure*}

\begin{figure*}[htbp]
\centering
\includegraphics[width=150mm]{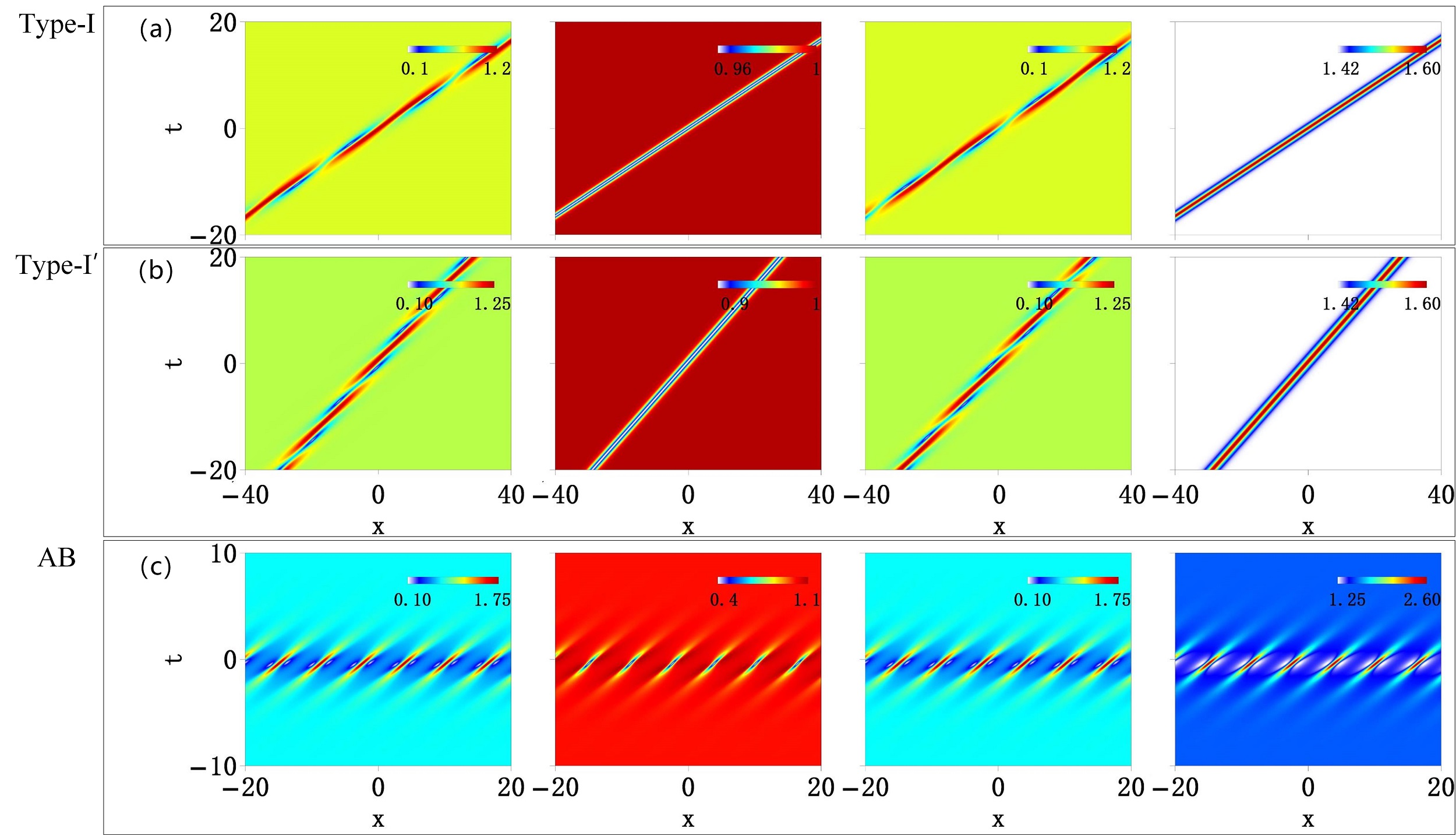}
\caption{Amplitude distributions of type-$I$ beating soliton (\ref{beating-s4}) (a) and type-$I'$ beating solitons (\ref{beating-s4}) (b), and vector bright-dark-bright Akhmediev breathers (c) in the defocusing regime. All these solutions correspond to the eigenvalue $\chi_{a,1}$.  From left to right, $|\psi_{}^{(1)}|$, $|\psi_{}^{(2)}|$, $|\psi_{}^{(3)}|$, and the total amplitude $\sqrt{|\psi_{}^{(1)}|^2+|\psi_{}^{(2)}|^2+|\psi_{}^{(3)}|^2}$. Parameters are $a=1$, $\beta=2$, $\alpha_1=0$ and $\gamma_1=1$.}
\label{f-f-beating2-b}
\end{figure*}

\begin{figure*}[htbp]
\centering
\includegraphics[width=150mm]{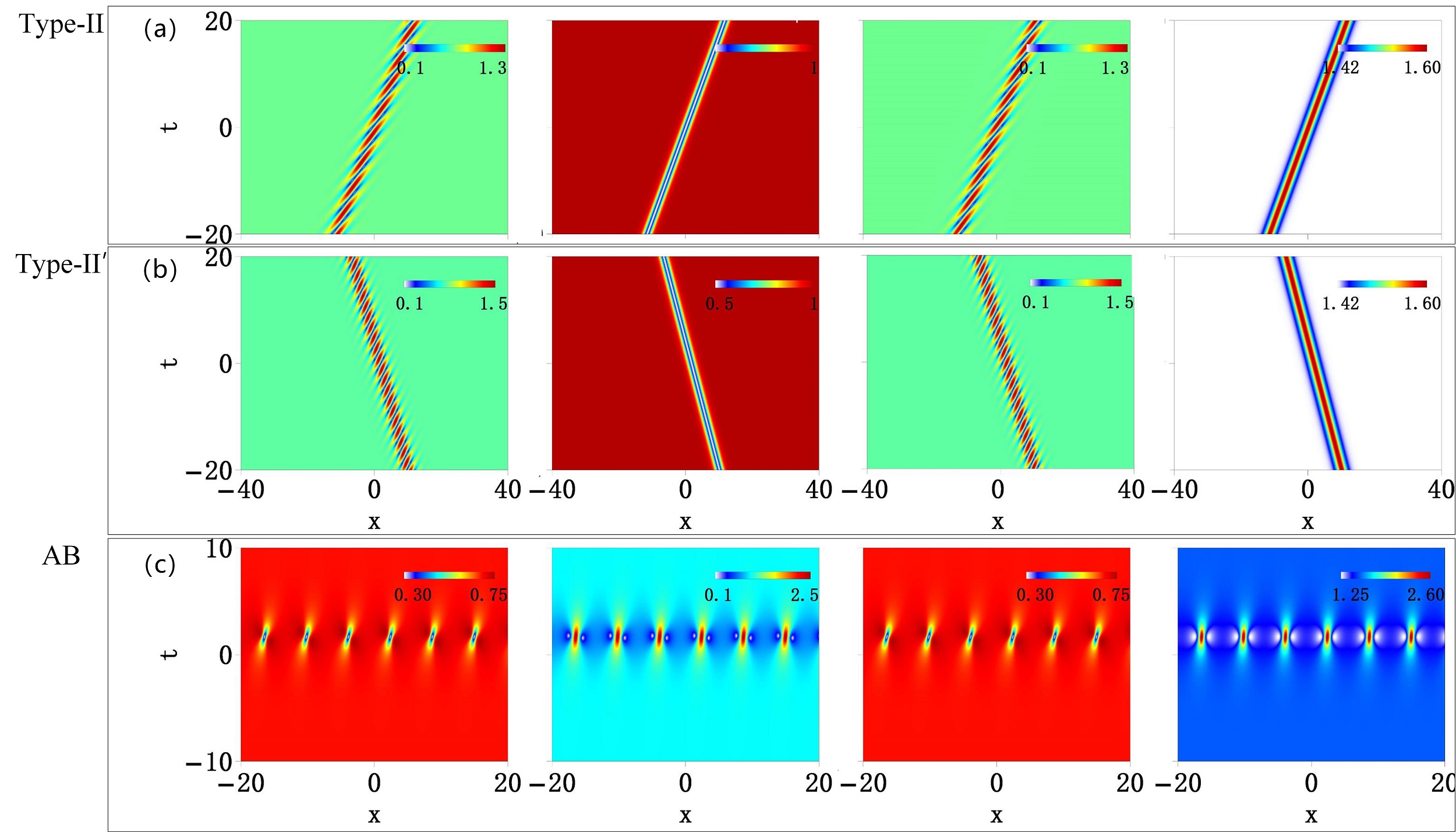}
\caption{Amplitude distributions of type-$II$ beating soliton (\ref{beating-s4}) (a) and type-$II'$ beating solitons (\ref{beating-s4}) (b), and vector dark-bright-dark Akhmediev breathers (c) in the focusing regime. All these solutions correspond to the eigenvalue $\chi_{a,2}$. From left to right, $|\psi_{}^{(1)}|$, $|\psi_{}^{(2)}|$, $|\psi_{}^{(3)}|$, and the total amplitude $\sqrt{|\psi_{}^{(1)}|^2+|\psi_{}^{(2)}|^2+|\psi_{}^{(3)}|^2}$. Parameters are $a=1$, $\beta=2$, $\alpha_1=0$ and $\gamma_1=1$.}
\label{f-f-beating2-c}
\end{figure*}

\subsection{The focusing regime}
We proceed to analyze the solutions (\ref{beating-s4}) in the focusing case $\delta=1$.
Just as the defocusing case, we consider the cases $\alpha=0$, $\gamma\neq0$.
We also plot the existence diagrams of the solutions.
Unlike the defocusing case, all four eigenvalues $\chi_{a,1}$, $\chi_{a,2}$, $\chi_{a,3}$, $\chi_{a,4}$ are valid in the focusing regime.
The analysis for each eigenvalue is more complex than the defocusing case.

Figure \ref{f-f-phase-1} presents existence diagrams of solutions on the $(\gamma, \beta)$ plane.
Beating solitons exist in two (cyan and gray) areas, but they are absent in the white regions.
The beating-soliton region also admits the vector ABs. In particular, we consider the beating solitons that can coexist with the vector ABs on the same plane wave background (see Section \ref{Sec-Second} about the higher-order solutions describing such coexistence and interaction).

Specifically, in the gray area, all four eigenvalues are valid for beating soliton formation; while only $\chi_{a,1}$ and $\chi_{a,2}$ are valid for AB formation. In particular, eigenvalues $\chi_{a,1}$ and $\chi_{a,2}$ play the same role in the AB formation.
Without loss of generality, we use $\chi_{a,1}$.
As shown in Fig. \ref{f-f-phase-1}, there are two types of beating solitons [$\psi^{(j)}(\chi_{a,1},\chi_{a,1})$, $\psi^{(j)}(\chi_{a,1},\chi_{b,1})$] and one type of ABs [$\psi^{(j)}(\chi_{a,1})$].

In contrast, in the cyan area, the role of $\chi_{a,1}$, $\chi_{a,3}$ (or $\chi_{a,2}$, $\chi_{a,4}$) in the formation of beating solitons and ABs is the same. Thus, we show in Fig. \ref{f-f-phase-1} (a) four types of beating solitons [$\psi^{(j)}(\chi_{a,1},\chi_{a,1})$, $\psi^{(j)}(\chi_{a,1},\chi_{b,1})$ for $\chi_{a,1}$; $\psi^{(j)}(\chi_{a,2},\chi_{a,2})$, $\psi^{(j)}(\chi_{a,2},\chi_{b,2})$ for $\chi_{a,2}$] and in Fig. \ref{f-f-phase-1} (b) two types of ABs [$\psi^{(j)}(\chi_{a,1})$, $\psi^{(j)}(\chi_{a,2})$].

Just as the defocusing case, the boundaries of the existence diagram can also be given explicitly.
Namely, the cyan area is limited by the curves
\begin{equation}\label{eq-efd1}
\beta_{c1}=\pm2a^2/\sqrt{4a^2-\gamma^2},
\end{equation}
and the gray area is limited by the curves
\begin{equation}\label{eq-efd2}
\begin{split}
&\beta_{c1}=\pm2a^2/\sqrt{4a^2-\gamma^2},~\textmd{where}~|\gamma|<\sqrt{2}, \\
&\beta_{c2}=\pm\gamma,~\textmd{where}~|\gamma|\geq\sqrt{2},\\
&\beta_{c3}=\pm\sqrt{\gamma^2-8a^2}.
\end{split}
\end{equation}
This condition coincides with that of vector ABs of two-component NLSEs in the focusing regime \cite{VAB-2022}.

Figure \ref{f-f-beating2-a} displays the individual and total component profiles for the gray area of the beating solitons and the ABs, which correspond to the point $(\gamma, \beta)=(1, 0.2)$ shown in Fig. \ref{f-f-phase-1}.
For the two types of beating solitons [$\psi^{(j)}(\chi_{a,1},\chi_{a,1})$, $\psi^{(j)}(\chi_{a,1},\chi_{b,1})$] shown in Fig. \ref{f-f-beating2-a} (a) and (b), oscillations occur only in the $\psi^{(1)}$ and $\psi^{(3)}$ components, while the $\psi^{(2)}$ component remains a non-oscillating dark soliton.
Interestingly, the two types of beating solitons have the same width and the opposite propagation velocity.
On the other hand, the coexisting AB [$\psi^{(j)}(\chi_{a,1})$] shown in Fig. \ref{f-f-beating2-a} (c) exhibits bright amplitude distribution.
We remark that this case is consistent with that of the two-component focusing NLSEs reported before \cite{VBE-2024}.

Let us then consider the amplitude profiles of the beating solitons and the ABs for the cyan area.
As shown in the existence diagram \ref{f-f-phase-1}, there are two families of solutions for different eigenvalues $\chi_{a,1}$ and $\chi_{a,2}$.   In Figs. \ref{f-f-beating2-b} and \ref{f-f-beating2-c}, we display these two different families of solutions, which correspond to the point $(\gamma, \beta)=(1, 2)$ shown in Fig. \ref{f-f-phase-1}.

As shown in Figs. \ref{f-f-beating2-b} and \ref{f-f-beating2-c}, all beating solutions show the beating-dark-beating soliton structure.
As the amplitudes of beating components $\psi^{(1)}$ and $\psi^{(3)}$ are complementary, the total amplitude shows a non-oscillating bright soliton.
An interesting finding is that the two types of vector ABs shown in Figs. \ref{f-f-beating2-b} and \ref{f-f-beating2-c} are different.
More specifically, the vector AB [$\psi^{(j)}(\chi_{a,1})$] in Fig. \ref{f-f-beating2-b} exhibits bright-dark-bright structure;
while the vector AB [$\psi^{(j)}(\chi_{a,2})$] in Fig. \ref{f-f-beating2-b} exhibits dark-bright-dark structure.
For either $\psi^{(j)}(\chi_{a,1})$ or $\psi^{(j)}(\chi_{a,2})$, we have $\psi^{(1)}=\psi^{(3)}$. This has been confirmed strictly in Section \ref{Sec-AB}.

\begin{figure*}[htbp]
\centering
\includegraphics[width=150mm]{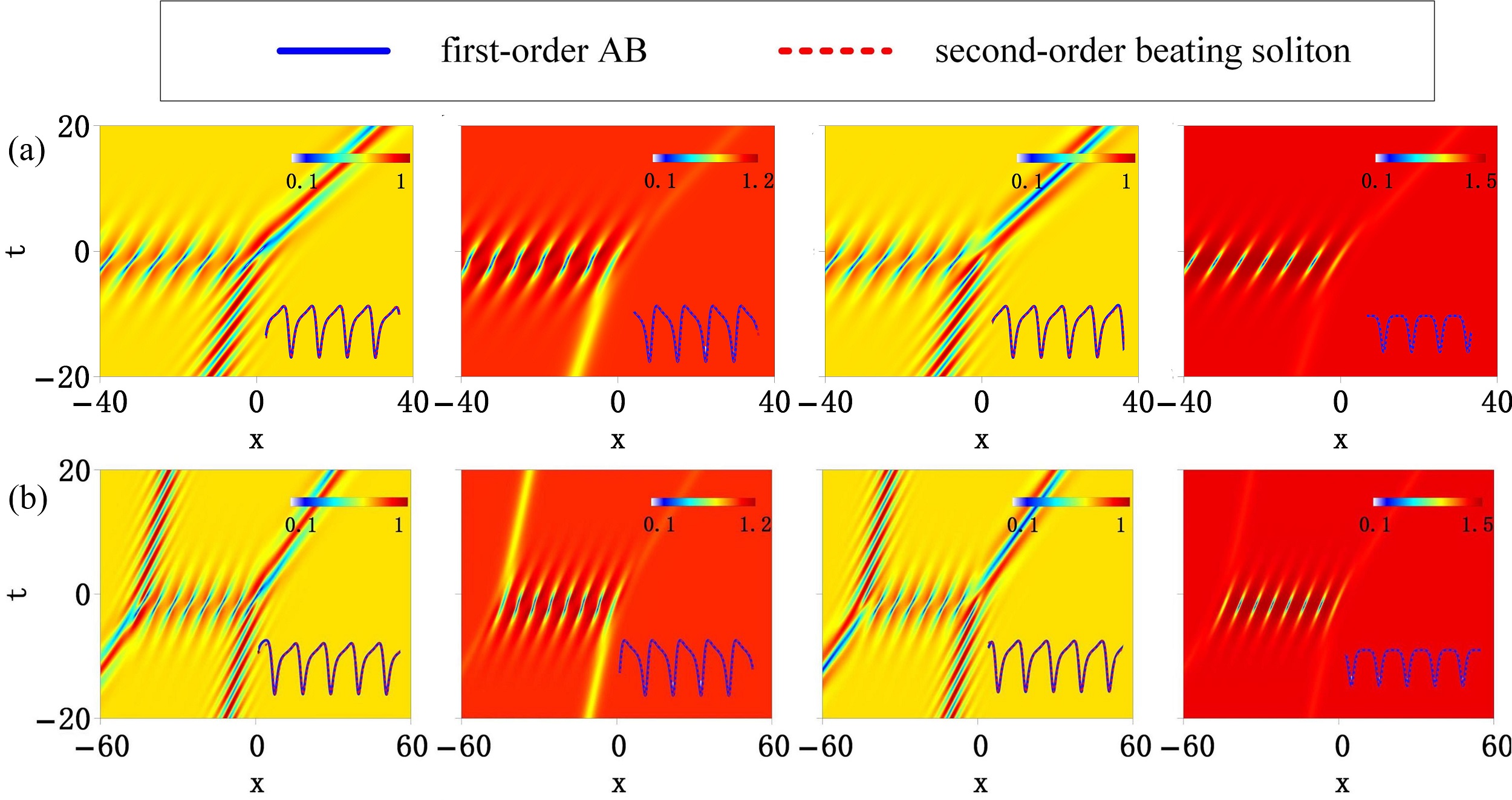}
\caption{Amplitude distributions of second-order beating soliton solutions $\psi^{(j)}[\chi_{a,1}(\gamma_1);\chi_{a,1}(\gamma_2)]$ formed by the nonlinear superposition of type-$I$ and type-$I'$ beating solitons in the defocusing regime under the conditions: (a) $\gamma_1=\gamma_2=1$; (b) $\gamma_1\cong\gamma_2$ ($\gamma_1=1$, $\gamma_2=1+10^{-8}$). The amplitude of the AB shown in Fig. \ref{f-d-beating2} (c) and the amplitude of the coexisting AB of second-order beating soliton solutions shown in (a) and (b) show great agreement.
Others are $a=1$, $\beta=0.2$, $\alpha_1=\alpha_2=0$.}
\label{f-d-second-1}
\end{figure*}

\begin{figure*}[htbp]
\centering
\includegraphics[width=150mm]{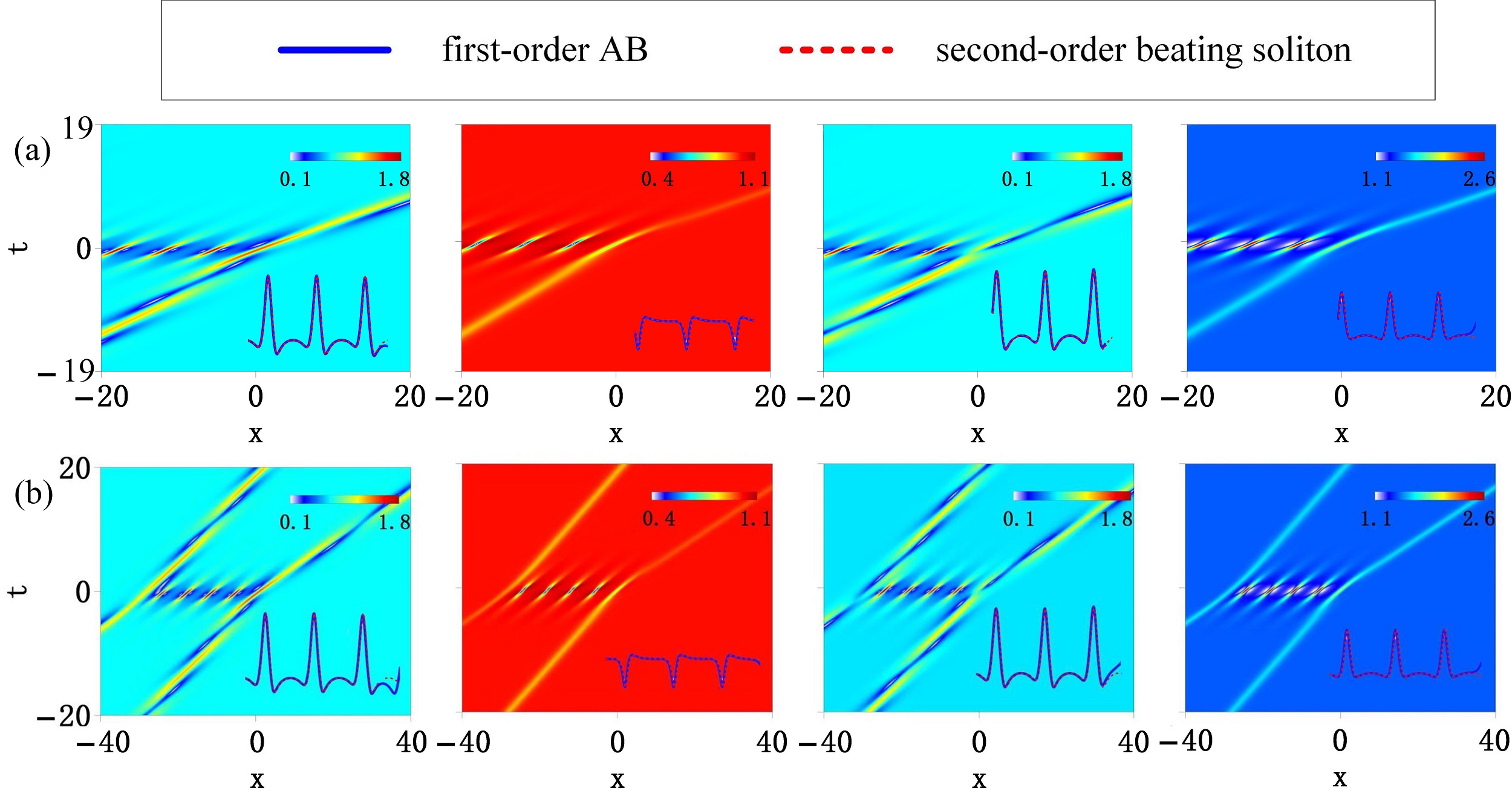}
\caption{Amplitude profiles of second-order solutions formed by the nonlinear superposition of type-$I$ and type-$I'$ beating solitons in the focusing regime under the conditions: (a) $\gamma_1=\gamma_2=1$; (b) $\gamma_1\cong\gamma_2$ ($\gamma_1=1$, $\gamma_2=1+10^{-8}$). The amplitude of the AB shown in Fig. \ref{f-f-beating2-b} (c) and the amplitude of the coexisting AB of second-order beating soliton solutions shown in (a) and (b) show great agreement.
Others are $a=1$, $\alpha_1=\alpha_2=0$, $\beta=2$.}
\label{f-f-second-1}
\end{figure*}

\begin{figure*}[htbp]
\centering
\includegraphics[width=150mm]{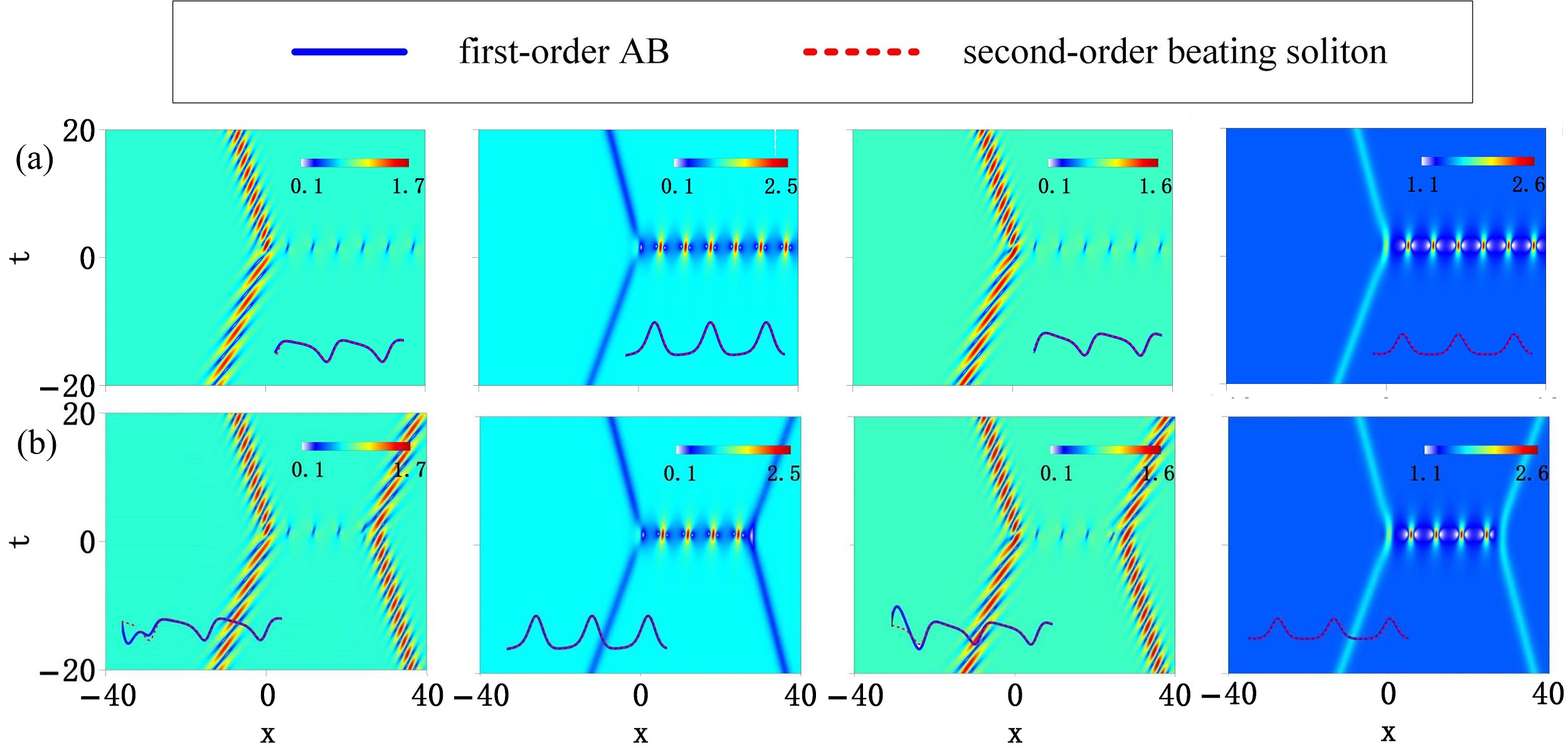}
\caption{Amplitude profiles of second-order solutions formed by the nonlinear superposition of type-$II$ and type-$II'$ beating solitons in the focusing regime under the conditions: (a) $\gamma_1=\gamma_2=1$; (b) $\gamma_1\cong\gamma_2$ ($\gamma_1=1$, $\gamma_2=1+10^{-8}$). The amplitude of the AB shown in Fig. \ref{f-f-beating2-b} (c) and the amplitude of the coexisting AB of second-order beating soliton solutions shown in (a) and (b) show great agreement.
Others are $a=1$, $\alpha_1=\alpha_2=0$, $\beta=2$.}
\label{f-f-second-2}
\end{figure*}

\section{Beating solitons constructed by rotation symmetry of vector solitons}\label{Sec-su}

As mentioned in Introduction \ref{Sec-Intro}, one can construct beating solitons by using the rotation symmetry of the vector solitons.
Here, we show that the beating soliton with nonzero relative wavenumber can also be constructed in this way only if certain types of vector solitons and certain rotation symmetry are considered.
Namely, we can obtain these beating solitons by performing a SU(2) transformation on dark-dark-bright solitons in both the defocusing and focusing regimes of the model (\ref{eq1}). On the other hand, we also consider the SU(2) and SU(3) transformations on dark-bright-bright solitons.
The case of SU(3) corresponds to the beating solitons with the same wavenumber (type ii) which we have shown in Appendix \ref{Sec-4}.

To show this point clearly, we first construct the exact solutions of dark-dark-bright solitons and dark-bright-bright solitons of the model (\ref{eq1}). The details are presented in Appendix \ref{Sec-5}.
We now focus our attentions on the SU(N) symmetry of the model (\ref{eq1}) given by
\begin{eqnarray}\label{eq-SU}
\left(
\begin{array}{cccc}
\psi^{(1)'}\\
\psi^{(2)'}\\
\psi^{(3)'}\\
\end{array}
\right)&=&\left(SU\right)\left(
\begin{array}{cccc}
\psi^{(1)}\\
\psi^{(2)}\\
\psi^{(3)}\\
\end{array}
\right),
\end{eqnarray}
where SU is a $3\times3$ constant matrix of form:
\begin{eqnarray}
SU&=&\left(
\begin{array}{cccc}\label{eq-SU0}
\kappa_1 & \kappa_2  & \kappa_3 \\
\kappa_4 & \kappa_5  & \kappa_6 \\
\kappa_7 & \kappa_8  & \kappa_9 \\
\end{array}
\right).
\end{eqnarray}
The SU matrix must meet the following conditions:
\begin{eqnarray}\label{S}
\begin{split}
(SU)(SU)^\dagger=I,~~~\det{(SU)}=1,\\
\end{split}
\end{eqnarray}
where $\dagger$ denotes the matrix transpose and complex conjugate, $I$ is
an identity matrix.

Equation (\ref{eq-SU}) means that if $(\psi^{(1)},\psi^{(2)},\psi^{(3)})^T$ is a solution to the system (\ref{eq1}), then $(\psi^{(1)'},\psi^{(2)'},\psi^{(3)'})^T$ obtained by the linear superposition is a new solution also satisfying (\ref{eq1}). Note that the total density profile is invariant upon rotation, i.e. $|\psi_1^{(1)}|^2+|\psi_1^{(2)}|^2+|\psi_1^{(3)}|^2$=
$|\psi_1^{(1)'}|^2+|\psi_1^{(2)'}|^2+|\psi_1^{(3)'}|^2$. In general, SU matrix (\ref{eq-SU0}) can be SU(3) or SU(2). However, to establish the link between the vector solitons and certain beating solitons we reported in this paper, we shall clarify the details as follows.

We first discuss how to construct the beating solitons with nonzero relative wavenumber via dark-dark-bright soliton under rotation symmetry.
To avoid the periodic amplitude background arising from the linear superposition between two dark-soliton components, we should apply SU(2) symmetry to the dark-dark-bright soliton. Thus, the elements of (\ref{eq-SU0}) should be either $\kappa_1=0$, $\kappa_7=0$  or $\kappa_2=0$, $\kappa_8=0$. Here, we use the condition $\kappa_2=0$, $\kappa_8=0$. Then, from (\ref{S}) we have
\begin{small}
\begin{eqnarray}\label{}
\left(
\begin{array}{cccc}
\kappa_1^2+\kappa_3^2 & \kappa_1\kappa_4+\kappa_3\kappa_6  & \kappa_1\kappa_7+\kappa_3\kappa_9 \\
\kappa_1\kappa_4+\kappa_3\kappa_6 & \kappa_4^2+\kappa_5^2+\kappa_6^2 & \kappa_4\kappa_7+\kappa_6\kappa_9\\
\kappa_1\kappa_7+\kappa_3\kappa_9 & \kappa_4\kappa_7+\kappa_6\kappa_9 & \kappa_7^2+\kappa_9^2\\
\end{array}
\right)=\left(
\begin{array}{cccc}
1 & 0  & 0 \\
0& 1 & 0\\
0 & 0 & 1\\
\end{array}
\right).
\end{eqnarray}
\end{small}
Let $\kappa_1=\kappa_3$, we have an explicit form of SU(2) as follows:
\begin{eqnarray}\label{eq-SU(2)}
SU(2)=\left(
\begin{array}{cccc}
\frac{\sqrt{2}}{2} & 0  & \frac{\sqrt{2}}{2} \\
0 & 1 & 0\\
\frac{\sqrt{2}}{2} & 0 & -\frac{\sqrt{2}}{2}\\
\end{array}
\right).
\end{eqnarray}
Substituting SU(2) matrix (\ref{eq-SU(2)}) into Eq. (\ref{eq-SU}), we have a new family beating solitons exhibiting beating-dark-beating structure with nonzero relative wavenumber. This solution coincides with the beating soliton solution (\ref{beating-s4}).

We then show how to construct the beating solitons with zero relative wavenumber via dark-bright-bright soliton under rotation symmetry.
If we perform the SU(2) transformation (\ref{eq-SU(2)}) for the dark-bright-bright soliton, we have the beating-bright-beating soliton.
However, if we perform the SU(3) transformation, we have the beating soliton exhibiting oscillation in each component. For a SU(3) matrix, we must have: $\kappa_1\neq0$, $\kappa_2\neq0$, and $\kappa_3\neq0$.
Here, we consider $\kappa_1=\kappa_2=\kappa_3$. From matrix (\ref{S}), we have a special case of SU(3) as follows:
\begin{eqnarray}
SU(3)&=&\frac{\sqrt{3}}{3}\left(
\begin{array}{cccc}\label{SU(3)}
1 & 1  & 1 \\
1 & \frac{-1+\sqrt{3}}{2} & \frac{-1-\sqrt{3}}{2}\\
1 & \frac{-1-\sqrt{3}}{2} & \frac{-1+\sqrt{3}}{2}\\
\end{array}
\right).
\end{eqnarray}
Using this SU(3) matrix, we have the beating solitons which coincide with the beating solitons with the same wavenumber (type ii) obtained by the Darboux transformation we have shown in Appendix \ref{Sec-4}.

\section{Coexistence and interaction
between beating solitons and vector ABs}\label{Sec-Second}

As shown in the two-component NLSEs \cite{VBE-2024,VSRB-2023}, the two types of beating solitons corresponding to the same eigenvalue can coexist with the scalar NLSE breathers only in the focusing regime. We show here that the results are special case when we consider the plane wave with the equal wavenumber.
In contrast, when we consider the plane wave with the nonzero relative wavenumber as Eq. (\ref{eqwnb}), we can obtain unique coexistence and interaction
between beating solitons and vector breathers in both focusing and defocusing regimes.

Below, we focus our attentions on the coexistence and interaction between the beating solitons and the vector ABs obtained in the previous sections.
They are described by the second-order solutions of beating solitons.
Similar to the solutions of the scalar NLSE \cite{SKM} and the two-component NLSEs \cite{VKMS-f2022,VKMS-df2023,VBE-2024},
the nonlinear superposition of fundamental vector solitons in the Manakov system can be constructed using next steps in the Darboux transformation (see Appendix \ref{Sec-2}).
In particular, we consider the second-order solutions with parameters $\gamma=\gamma_1$ and $\gamma=\gamma_2$.
For either the defocusing or focusing regime, we consider two cases: i) $\gamma_1=\gamma_2$; ii) $\gamma_1\cong\gamma_2$.

Figure \ref{f-d-second-1} (a) shows, in the defocusig regime, the corresponding amplitude profiles of second-order solutions $\psi^{(j)}[\chi_{a,1}(\gamma_1);\chi_{a,1}(\gamma_2)]$ with $\gamma_1=\gamma_2$.
As can be seen from the figure, a typical Y-shaped structure exhibiting the coexistence of three different types of localized waves shown in Fig. \ref{f-d-beating2} is observed.
Namely, a type-$I'$ beating soliton [shown in Fig. \ref{f-d-beating2} (b)] collides with a vector dark AB [shown in Fig. \ref{f-d-beating2} (c)] (around $t=0$), and then they form a type-$I$ beating soliton [shown in Fig. \ref{f-d-beating2} (a)]. It is evident that the ABs produced by the coexistence is consistent with the AB shown in Fig. \ref{f-d-beating2} (c), as indicated in the inset of Fig. \ref{f-d-second-1}.

Let us consider the case $\gamma_1\cong\gamma_2$.
As shown in Fig. \ref{f-d-second-1} (b) the second-order solutions $\psi^{(j)}[\chi_{a,1}(\gamma_1);\chi_{a,1}(\gamma_2)]$ exhibits
the interaction between two (type-$I'$ and type-$I$) beating solitons.
The soliton collision is more complex than the case shown in Fig. \ref{f-d-second-1} (a). As can be seen, these two solitons collide with each other and form an intermediate breather structure. The latter is the vector dark AB. It is demonstrated that such dark AB has the same amplitude with that shown in Fig. \ref{f-d-beating2} (c). The comparison of the amplitude profiles is shown in the insets of the figure.
We point out that the collision is elastic which has been proved by the asymptotic analysis of the two solitons. %The details are tedious.
Since the procedure of  the asymptotic analysis is the same as shown in Ref. \cite{VBE-2024}, thus we omit the details here.

Next, we will consider the second-order solutions in the focusing regime.
Just as the defocusing case, the solutions describe the
interaction between the beating solitons and their coexistence with AB.
In particular, for the case of beating solitons and ABs in the gray region shown in Fig. \ref{f-f-phase-1},
only second-order solutions $\psi^{(j)}[\chi_{a,1}(\gamma_1);\chi_{a,1}(\gamma_2)]$ are allowed.
The soliton-collision patterns describe the coexistence of beating solitons and ABs shown in Fig. \ref{f-f-beating2-a} [not shown here], which are similar with those of the defocusing case shown in Fig. \ref{f-d-second-1}.

Notably, the cyan region reveals that type-$I$ (type-$II$) and type-$I'$ (type-$II'$) beating soliton collisions exhibit a more interesting phenomenon. We first consider the nonlinear
superpositions of type-$I$ and type-$I'$ beating solitons. Figures \ref{f-f-second-1} (a) and \ref{f-f-second-1} (b)
show the amplitude profiles of such second-order solutions. Similar to the soliton collisions shown in Fig. \ref{f-d-second-1}, these collisions can generate ABs. Figure \ref{f-f-second-1} (a) shows examples of the amplitude profiles with $\gamma_1=\gamma_2$. There are three fundamental modes (type-$I$ and type-$I'$ beating solitons and a bright-dark-bright AB) that correspond to the same eigenvalue $\chi_{a,1}$. The resulting patterns is the fusion of two fundamental modes. Figure \ref{f-f-second-1}(b) shows examples of the amplitude profiles with $\gamma_1\cong\gamma_2$. The bright-dark-bright breather structures are generated during
the collision.

Let us now consider the case of second-order solutions correspond to the eigenvalue $\chi_{a,2}$. Just as the case of $\chi_{a,1}$ shown in Fig. \ref{f-f-second-1}, we consider the two (type-$II$ and type-$II'$) beating solitons with $\gamma_1=\gamma_2$ and $\gamma_1\cong\gamma_2$.
Figure \ref{f-f-second-2} (a) presents the amplitude distributions for the case $\gamma_1=\gamma_2$. The pattern shown in Fig. \ref{f-f-second-2} (a)
exhibits the coexistence of two beating solitons and a dark-bright-dark AB. This is in sharp contrast to the case shown in Fig. \ref{f-f-second-1}, where the AB exhibits a bright-dark-bright breather structure.
As the beating solitons have opposite group velocities, the pattern can be considered as a
reflection of a beating soliton from an AB around $t=0$.
Figure \ref{f-f-second-2} (b) shows examples of the two-beating soliton collision with $\gamma_1\cong\gamma_2$ where the intermediate breather structure is a dark-bright-dark AB.

\begin{figure*}[htbp]
	\centering
	\includegraphics[width=154mm]{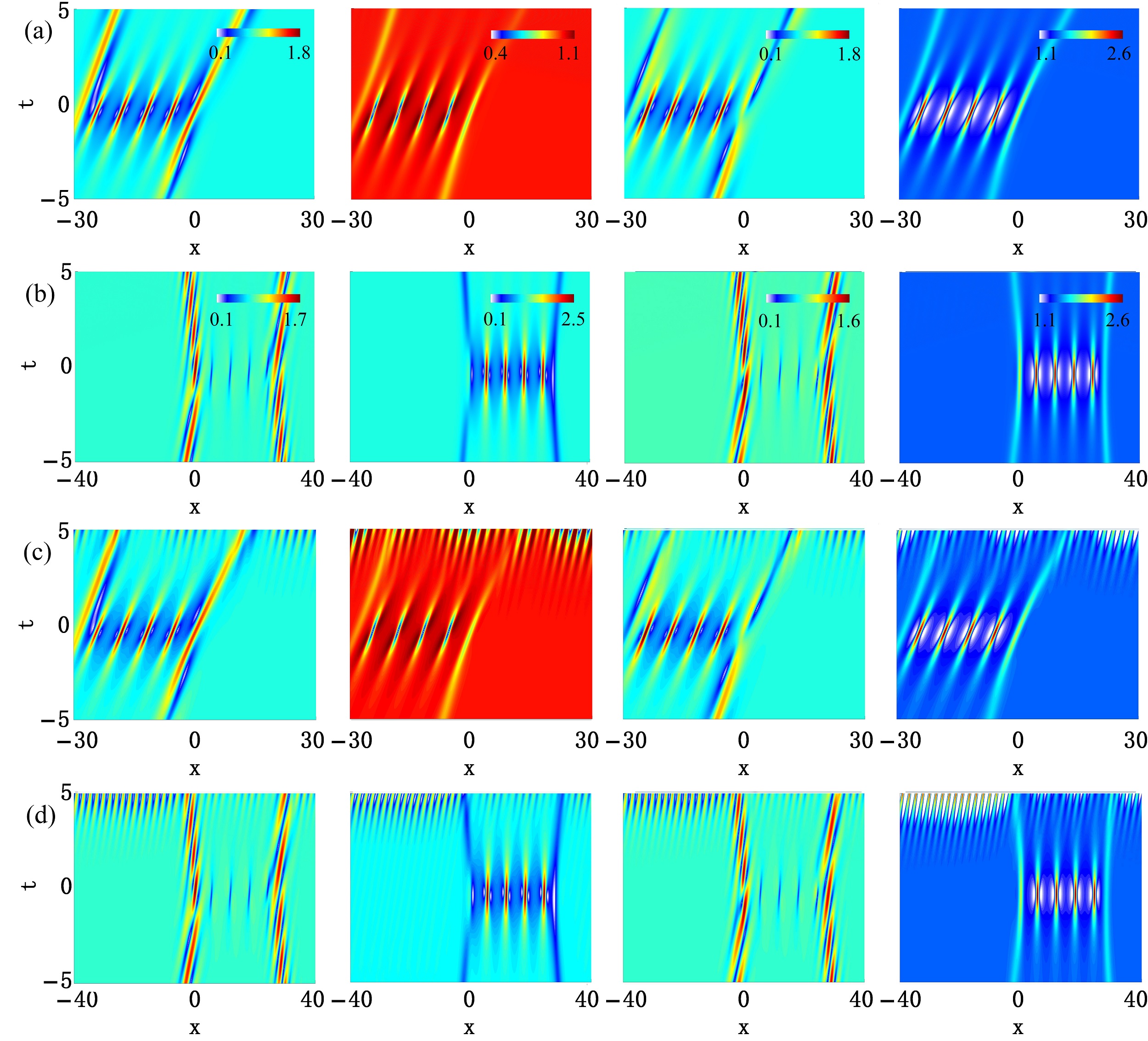}
	\caption{Numerical simulation of second-order solutions shown in Figs. \ref{f-f-second-1} (a) and \ref{f-f-second-2} (a) starting from two initial conditions: (a-b) exact solution $\psi^{(j)}(x,t=-5)$; (c-d) exact solution $\psi^{(j)}(x,t=-5)$ perturbed by the white noise of a strength of $10^{-5}$. As can be seen, all solutions are robust against the unavoidable numerical and white noises.}
	\label{f-num1}
\end{figure*}

\section{Numerical simulations}\label{Sec-Numerical}

Finally, let us confirm the validity of our exact solutions by using the numerical simulations. We perform numerical simulations by solving Eq. (\ref{eq1}) with the split-step Fourier method.
The initial conditions we use here are extracted from the exact solutions $\psi^{(j)}(x,t=-5)$ at $t=-5$. Moreover, we consider the initial condition perturbed by the white noise of a strength of $10^{-5}$ in order to show better the robustness of the solutions.
To give an example, we consider the numerical simulations of the two-beating soliton collision shown in Figs. \ref{f-f-second-1} (a) and \ref{f-f-second-2} (a). The results of numerical simulations are shown in Fig. \ref{f-num1}.
Figures \ref{f-num1}(a) and \ref{f-num1}(b) show the numerical evolution starting with the exact initial condition $\psi^{(j)}(x,t=-5)$ without weak white noises. As can be seen from the figure, all complex higher-order exact solution describing the wave collisions are perfectly reproduced.
Figures \ref{f-num1}(c) and \ref{f-num1}(d) show the numerical evolution starting with the exact initial condition $\psi^{(j)}(x,t=-5)$ perturbed by the weak white noise of a strength of $10^{-5}$.
The numerical simulation follows the exact solution until $t \approx 5$ when the spontaneous modulation instability pattern emerges on the plane wave. The latter is unavoidable for the plane wave background.
Generally, all our numerical simulations show good agreement with the exact solutions.

\section{Conclusions}\label{Sec-Conc}
We have identified a new family of beating soliton solutions with nonzero relative wavenumber of three-component NLSEs in both the focusing and defocusing regimes. Based on the exact solutions,
we have presented the existence diagrams of these beating solitons. By using the existence diagrams, we have found unique coexistence and interaction between beating solitons and vector Akhmediev breathers. It is demonstrated that the coexisting breathers turn out to be the two-component Akhmediev breathers with nonzero relative wavenumber.
The validity of our exact solutions has been confirmed by numerical simulations.
Our results could have direct impact in the experimental side where three-component solitons have been observed in Bose-Einstein condensates \cite{three-2020}.

\section*{ACKNOWLEDGEMENTS}
The work is supported by the NSFC (Grants No. 12175178, and No. 12247103), Shaanxi Fundamental Science Research Project for Mathematics and Physics (Grant No. 22JSY016), and 2023 Graduate Innovation Program of NorthwestUniversity No.CX2024137.

\begin{appendix}

\section{Types of beating solitons}\label{Sec-0}
To construct the beating soliton solutions, we first represent Eq. (\ref{eq1}) in the form of
two linear equations with $4\times4$ matrix operators:
\begin{eqnarray}
\bf{\Psi}_x=\textbf{U}\Psi,~~~\bf{\Psi}_t=\textbf{V}\Psi,\label{Lax-pair}
\end{eqnarray}
where
\begin{small}
\begin{eqnarray}
\begin{split}\label{UV1}
&~~~~~~~~~~~~~~~\mathbf{U}=i\left(\frac{\lambda}{2}\left(\bm\sigma_4+\mathbf{I}\right)+\mathbf{Q}\right),\\
&\mathbf{V}=i\left(\frac{\lambda^2}{4}\left(\bm\sigma_4+\mathbf{I}\right)+
\frac{\lambda}{2}\mathbf{Q}-\frac{1}{2}\bm\sigma_4(\mathbf{Q}^2+i\mathbf{Q}_x)+A\mathbf{I}\right)
\end{split}
\end{eqnarray}
\end{small}
Where
\begin{eqnarray}
\mathbf{Q}&=&\left(
\begin{array}{ccccc}
0& \delta\psi^{(1)*}& \delta\psi^{(2)*}& \delta\psi^{(3)*}\\
\psi^{(1)}& 0& 0& 0\\
\psi^{(2)}& 0& 0& 0\\
\psi^{(3)}& 0& 0& 0\\
\end{array}
\right),\\
\bm{\sigma}_4&=&\left(
\begin{array}{ccccc}
1& 0& 0& 0\\
0 & -1& 0& 0\\
0 & 0& -1& 0\\
0 & 0& 0& -1\\
\end{array}
\right).
\end{eqnarray}
Here, the vector function $\bm{\Psi}$=$\left({R, S, W, X}\right)^\textsf{T}$ ($\textsf{T}$ means a matrix transpose), $*$ denotes complex conjugate, $\mathbf{I}$ is an identity matrix, $\lambda$ is the spectral parameter, and $A$ is a real parameter.
The system of Manakov equations (\ref{eq1}) follows from the compatibility condition
\begin{eqnarray}
\mathbf{U}_t-\mathbf{V}_x+[\mathbf{U}, \mathbf{V}]=0.
\end{eqnarray}
In order obtain the beating soliton solution, we shall start with the plane wave solution $\psi_{0}^{(j)}$ as the seed solution, which is given by:
\begin{equation}
\begin{split}
\psi_{0}^{(j)}=a_j\exp{\{i\theta_j\}},~\theta_j=\beta_j x + \delta(a_1^2+a_2^2+a_3^2- \frac{1}{2}\beta_j^2 )t,\\
\end{split}\label{eqpw}
\end{equation}
where $a_j$, $\beta_j$ are the amplitudes and wavenumbers of the $\psi^{(j)}$ component, respectively.

In our previous works \cite{VKMS-f2022,VKMS-df2023,VBE-2024}, beating solitons exist only when the vector plane wave components have the same wavenumber. Interestingly, we find that there are still beating solitons in the three-component NLSEs with nonzero relative wavenumber:
\begin{equation}
\beta_1=\beta_3=\beta,~\beta_2=0. \label{eqwnb1}
\end{equation}
Submitting (\ref{eqpw}) into the Lax pair with $A=a_1^2+a_2^2+a_3^2$ and using a diagonal matrix $\mathbf{S}$=diag$(1,e^{-i\theta_1},e^{-i\theta_2},e^{-i\theta_3})$, we can rewrite the Lax pair as:
%The Lax pair (\ref{UV1}) can be written in the following form:
\begin{eqnarray}\label{UV2}
\begin{small}
\begin{split}
&\tilde{\mathbf{U}}=i\left(
\begin{array}{ccccc}
\lambda & \delta a_1 & \delta a_2 & \delta a_3\\
a_1 & -\beta_1 & 0 &0\\
a_2 & 0 & -\beta_2 &0\\
a_3 & 0 & 0 &-\beta_3\\
\end{array}
\right),
&\tilde{\mathbf{V}}=-\frac{i}{2}\tilde{\mathbf{U}}^2+i(1-\delta)A\mathbf{I},
\end{split}
\end{small}
\end{eqnarray}
The linear eigenvalue problem in terms of the transformed Lax pair (\ref{UV1}) is given by
\begin{eqnarray}\label{det}
\det(\tilde{\mathbf{U}}-i\chi)=0.
\end{eqnarray}
Eq. (\ref{det}) admits four eigenvalues $\chi_{l}$ ($l=a,b,c,d$) which
can be transformed into the spectral parameters
\begin{equation}
\label{eqlambda} \lambda=\chi_{l}-\delta\frac{a_1^2}{\chi_{l}+\beta_1}-\delta\frac{a_2^2}{\chi_{l}+\beta_2}-\delta\frac{a_3^2}{\chi_{l}+\beta_3}.
\end{equation}
The first of them can be simply shifted in the complex
plane to give the second one,
\begin{equation}
\chi_{b}=\chi_{a}+i\alpha+\gamma,\label{eqchi-b}
\end{equation}
where $\alpha$ and $\gamma$ are two important real parameters (we will show the meaning below).
Now, using (\ref{eqlambda}), we obtain the equation for finding the eigenvalues:
\begin{equation}
1+\sum_{i=1}^{3}\delta\frac{a_j^2}{(\chi_a+\beta_j)(\chi_{b}+\beta_j)}=0.\label{eqchi}
\end{equation}
Moreover, to simplify our analysis we take here
\begin{equation}
a_1=a_3=a/\sqrt{2},~a_2=a. \label{eqpwa1}
\end{equation}
Equation (\ref{eqchi}) has solutions
\begin{eqnarray}\label{b-eqchi1234}
\begin{split}		
\chi_{a,1}&=\frac{1}{2}i\left(-\alpha+i(\beta+\gamma)-\sqrt{\mu-2i\sqrt{\eta}}\right),\\
\chi_{a,2}&=\frac{1}{2}i\left(-\alpha+i(\beta+\gamma)+\sqrt{\mu-2i\sqrt{\eta}}\right),\\
\chi_{a,3}&=\frac{1}{2}i\left(-\alpha-i(\beta+\gamma)+\sqrt{\mu+2i\sqrt{\eta}}\right),\\
\chi_{a,4}&=\frac{1}{2}i\left(-\alpha-i(\beta+\gamma)-\sqrt{\mu+2i\sqrt{\eta}}\right).
\end{split}
\end{eqnarray}
where
\begin{eqnarray}
\begin{split}		
&\mu=\alpha^2-\beta^2-2i\alpha\gamma-\gamma^2+4a^2\delta,\\
&\eta=\alpha^2\beta^2-2i\alpha\gamma\beta^2-\beta^2\gamma^2+4a^2\beta^2\delta-4a^4\delta^2.
\end{split}
\end{eqnarray}
For each of $\chi_{a}$, we have the corresponding spectral parameter $\lambda(\chi_{a})$ given by (\ref{eqlambda}). Substituting $\lambda(\chi_{a})$ into Eq.(\ref{det}), we numerically obtain the third eigenvalue $\chi_{c}$. The fourth eigenvalue is given by
\begin{equation}	
\chi_{d}=-\beta.
\end{equation}
We further diagonalize the matrixes $\tilde{\mathbf{U}}$ and $\tilde{\mathbf{V}}$. Namely, we have
\begin{equation}	
\varphi_{x}=\tilde{\mathbf{U}}\varphi,~~~~\varphi_{t}=\tilde{\mathbf{V}}\varphi,
~~~~\varphi=\mathbf{H}^{-1}\mathbf{S}\bf{\Psi},\label{UV3}
\end{equation}
where the transformation matrix $\mathbf{H}$ is
\begin{eqnarray}\label{EqH3-2}
\begin{split}
&\mathbf{H}=\left(
\begin{array}{cccc}
1 & \frac{\beta_1+\chi_{n,b}}{a_1} & \frac{\beta_2+\chi_{n,c}}{a_2}  & \frac{\beta_3+\chi_{n,d}}{a_3}\\
\frac{a_1}{\beta_1+\chi_{n,a}} & 1 & \frac{(\beta_2+\chi_{n,c})}{(\beta_1+\chi_{n,c})} & -1\\
\frac{a_2}{\beta_2+\chi_{n,a}} & \frac{(\beta_1+\chi_{n,b})}{(\beta_2+\chi_{n,b})} & 1 & 0\\
\frac{a_3}{\beta_3+\chi_{n,a}} & \frac{(\beta_1+\chi_{n,b})}{(\beta_3+\chi_{n,b})} & \frac{(\beta_2+\chi_{n,c})}{(\beta_3+\chi_{n,c})} & 1\\
\end{array}
\right).\\
\end{split}
\end{eqnarray}
Solving Eq. (\ref{UV3}) , we have
\begin{equation}	
\varphi_{n,l}=c_{n,l}\exp{\left[i\left(\chi_{l}x+\frac{1}{2}(\chi_{l}^2+2(1-\delta)A) t\right)\right]}.
\end{equation}
Here $c_{n,l}$ are arbitrary real constants.
Finally, the eigenfunctions $\bf{\Psi}_1$=$(R_{1},S_{1},W_{1},X_{1})$ ($n=1$) of the Lax pair (\ref{Lax-pair}) are given by
\begin{eqnarray}\label{}
\begin{split}
\bf{\Psi}_1=\left(
\begin{array}{cccc}
R_{1}\\
S_{1}\\
W_{1}\\
X_{1}\\
\end{array}
\right)=\mathbf{S}.\mathbf{H}.\left(
\begin{array}{cccc}
\varphi_{1,a}\\
\varphi_{1,b}\\
\varphi_{1,c}\\
\varphi_{1,d}\\
\end{array}
\right).
\end{split}
\end{eqnarray}
Substituting the eigenfunctions into (\ref{eqdt1}), we have the first-order beating soliton solutions.
Beating solitons can be classified using the eigenvalues ($\chi_{a,1}$, $\chi_{a,2}$, $\chi_{a,3}$, $\chi_{a,4}$) and the coefficients $\{c_{1,a}, c_{1,b}, c_{1,c}, c_{1,d}\}$. Such classification is presented in Table \ref{Table3}. As we have found, there are twelve types of beating solitons in the focusing regime and eight types in the defocusing regime.

It should be pointed out that
when $\beta_1=\beta_3=0$ in (\ref{eqwnb1}), the transformation matrix $\mathbf{H}$ (\ref{EqH3-2}) can be rewritten as:
\begin{eqnarray}\label{EqH3-0}
\begin{split}
&\mathbf{H}=\left(
\begin{array}{cccc}
1 & \frac{\beta_1+\chi_{n,b}}{a_1} & \frac{\beta_2+\chi_{n,c}}{a_2}  & \frac{\beta_3+\chi_{n,d}}{a_3}\\
\frac{a_1}{\beta_1+\chi_{n,a}} & 1 & 0 & -1\\
\frac{a_2}{\beta_2+\chi_{n,a}} & \frac{(\beta_1+\chi_{n,b})}{(\beta_2+\chi_{n,b})} & 1 & 0\\
\frac{a_3}{\beta_3+\chi_{n,a}} & \frac{(\beta_1+\chi_{n,b})}{(\beta_3+\chi_{n,b})} & -1 & 1\\
\end{array}
\right).\\
\end{split}
\end{eqnarray}
Using matrix $\mathbf{H}$ (\ref{EqH3-0}), we then obtain a special family of beating solitons with zero relative wavenumber.  Such beating soliton also exhibits the beating-dark-beating structure. However, it can only coexist with the scalar ABs in the focusing regime, just like the case of two-component NLSEs we have studied before in Ref. \cite{VBE-2024}.

\begin{table*}[htb]
\begin{center}
\renewcommand{\arraystretch}{1.6}
\begin{tabular}{|m{1cm}<{\centering}|m{3.2cm}<{\centering}|m{2.5cm}<{\centering}|m{5cm}<{\centering}|m{4cm}<{\centering}|}
\hline
&\multicolumn{4}{c|}{$\beta_1=\beta_3=\beta,\beta_2=0$} \\
\hline
&\multicolumn{2}{c|} {Fundamental Beating Solitons} & Coefficients $\{c_{1,a},c_{1,b},c_{1,c},c_{1,d}\}$ & Parameter ($\chi_k$) \\
\hline
\multirow{12}*{$\delta=1$}
&Type-$I$ &$\psi^{(j)}(\chi_{a,1})$ & $\{c_{1,a},0,0,c_{1,d}\}$ &$\chi_k=\chi_{a,1}$  \\
\cline{2-5}
&Type-$I'$ &$\psi^{(j)}(\chi_{a,1})$ & $\{0,c_{1,b},0,c_{1,d}\}$ &$\chi_k=\chi_{b,1}$  \\
\cline{2-5}
&Type-$I''$ &$\psi^{(j)}(\chi_{a,1})$ & $\{0,0,c_{1,c},c_{1,d}\}$ &$\chi_k=\chi_{c,1}$  \\
\cline{2-5}
&Type-$II$ &$\psi^{(j)}(\chi_{a,2})$ & $\{c_{1,a},0,0,c_{1,d}\}$ &$\chi_k=\chi_{a,2}$  \\
\cline{2-5}
&Type-$II'$ &$\psi^{(j)}(\chi_{a,2})$ & $\{0,c_{1,b},0,c_{1,d}\}$ &$\chi_k=\chi_{b,2}$  \\
\cline{2-5}
&Type-$II''$ &$\psi^{(j)}(\chi_{a,2})$ & $\{0,0,c_{1,c},c_{1,d}\}$ &$\chi_k=\chi_{c,2}$  \\
\cline{2-5}
&Type-$III$ &$\psi^{(j)}(\chi_{a,3})$ & $\{c_{1,a},0,0,c_{1,d}\}$ &$\chi_k=\chi_{a,3}$  \\
\cline{2-5}
&Type-$III'$ &$\psi^{(j)}(\chi_{a,3})$ & $\{0,c_{1,b},0,c_{1,d}\}$ &$\chi_k=\chi_{b,3}$  \\
\cline{2-5}
&Type-$III''$ &$\psi^{(j)}(\chi_{a,3})$ & $\{0,0,c_{1,c},c_{1,d}\}$ &$\chi_k=\chi_{c,3}$  \\
\cline{2-5}
&Type-$IV$ &$\psi^{(j)}(\chi_{a,4})$ & $\{c_{1,a},0,0,c_{1,d}\}$ &$\chi_k=\chi_{a,4}$  \\
\cline{2-5}
&Type-$IV'$ &$\psi^{(j)}(\chi_{a,4})$ & $\{0,c_{1,b},0,c_{1,d}\}$ &$\chi_k=\chi_{b,4}$  \\
\cline{2-5}
&Type-$IV''$ &$\psi^{(j)}(\chi_{a,4})$ & $\{0,0,c_{1,c},c_{1,d}\}$ &$\chi_k=\chi_{c,4}$  \\
\hline
\multirow{12}*{$\delta=-1$}
&Type-$I'$ &$\psi^{(j)}(\chi_{a,1})$ & $\{0,c_{1,b},0,c_{1,d}\}$ &$\chi_k=\chi_{b,1}$  \\
\cline{2-5}
&Type-$I$ or Type-$I''$  &$\psi^{(j)}(\chi_{a,1})$ & $\{c_{1,a},0,0,c_{1,d}\}$ or $\{0,0,c_{1,c},c_{1,d}\}$ &$\chi_k=\chi_{a,1}$ or $\chi_k=\chi_{c,1}$  \\
\cline{2-5}
&Type-$II$ &$\psi^{(j)}(\chi_{a,2})$ & $\{c_{1,a},0,0,c_{1,d}\}$ &$\chi_k=\chi_{a,2}$  \\
\cline{2-5}
&Type-$II'$ or Type-$II''$  &$\psi^{(j)}(\chi_{a,2})$ & $\{0,c_{1,b},0,c_{1,d}\}$ or $\{0,0,c_{1,c},c_{1,d}\}$
&$\chi_k=\chi_{b,2}$ or $\chi_k=\chi_{c,2}$  \\
\cline{2-5}
&Type-$III'$ &$\psi^{(j)}(\chi_{a,3})$ & $\{0,c_{1,b},0,c_{1,d}\}$ &$\chi_k=\chi_{b,3}$  \\
\cline{2-5}
&Type-$III''$ &$\psi^{(j)}(\chi_{a,3})$ & $\{0,0,c_{1,c},c_{1,d}\}$ &$\chi_k=\chi_{c,3}$  \\
\cline{2-5}
&Type-$IV$ &$\psi^{(j)}(\chi_{a,4})$ & $\{c_{1,a},0,0,c_{1,d}\}$ &$\chi_k=\chi_{a,4}$  \\
\cline{2-5}
&Type-$IV''$ &$\psi^{(j)}(\chi_{a,4})$ & $\{0,0,c_{1,c},c_{1,d}\}$ &$\chi_k=\chi_{c,4}$  \\
\hline
\end{tabular}
\caption{A summary for different types of beating solitons (\ref{beating-s4}) $\psi^{(j)}(\chi_{a,1})$ and $\psi^{(j)}(\chi_{a,2})$ in both focusing and defocusing regimes ($\delta=\pm1$) with different combinations of coefficients $\{c_{1,a},c_{1,b},c_{1,c}\}$ and with different parameters $\chi_k$ used in solutions (\ref{beating-s4}).}\label{Table3}
\end{center}
\end{table*}

\section{Fundamental beating solitons with $\beta_1=\beta_2=\beta_3$}\label{Sec-4}
%This section, we consider the case with the same wavenumber:
Here, we consider the case of three-component beating solitons with the same wavenumber:
\begin{equation}
\beta_1=\beta_2=\beta_3=\beta. \label{4-eqwn}
\end{equation}
Note that in this case, beating soliton can only coexist with scalar NLSE breathers.
The eigenvalue equation (\ref{eqchi}) has solutions
\begin{eqnarray}\label{4-eqchi12}
\begin{split}		
\chi_{a,1}&=\frac{1}{2}\left(-i\alpha-2\beta-\gamma-\sqrt{-(\alpha-i\gamma)^2-4A\delta}\right),\\
\chi_{a,2}&=\frac{1}{2}\left(-i\alpha-2\beta-\gamma+\sqrt{-(\alpha-i\gamma)^2-4A\delta}\right),
\end{split}
\end{eqnarray}
where $A=a_1 ^2+a_2^2+a_3^2$.
For each of $\chi_{a}$, we have the corresponding spectral parameter $\lambda(\chi_{a})$ given by (\ref{eqlambda}). Substituting $\lambda(\chi_{a})$ into Eq.(\ref{det}), we obtain the third and fourth eigenvalues
\begin{eqnarray}\label{4-eqchiabcd}
\begin{split}		
\chi_{c}=-\beta,~~~\chi_{d}=-\beta.
\end{split}
\end{eqnarray}
Here, the transformation matrix $\mathbf{H}$ is
\begin{eqnarray}\label{4-EqH3-1}
\begin{split}
&\mathbf{H}=\left(
\begin{array}{cccc}
1 & \frac{\beta_1+\chi_{n,b}}{a_1} & \frac{\beta_2+\chi_{n,c}}{a_2}  & \frac{\beta_3+\chi_{n,d}}{a_3}\\
\frac{a_1}{\beta_1+\chi_{n,a}} & 1 & \rho_1 & \rho_2\\
\frac{a_2}{\beta_2+\chi_{n,a}} & \frac{(\beta_1+\chi_{n,b})}{(\beta_2+\chi_{n,b})} & 1 & (-1-\rho_2)\\
\frac{a_3}{\beta_3+\chi_{n,a}} & \frac{(\beta_1+\chi_{n,b})}{(\beta_3+\chi_{n,b})} & (-1-\rho_1) & 1\\
\end{array}
\right).\\
\end{split}
\end{eqnarray}
Here, $\rho_1$ and $\rho_2$ are arbitrary constants. Choice of these two parameters yields different beating solitons which can be classified into the i) and ii) types:.

i) When $\rho_2=-1$, $\rho_1=0$,
the transformation matrix $\mathbf{H}$ (\ref{4-EqH3-1}) reduces to (\ref{EqH3-0}).
The resulting beating soliton exhibits the beating-dark-beating structure (\ref{EqH3-0}). We also find that when $\rho_1=-1$, $\rho_2=0$, the corresponding beating solitons are the same as the case of $\rho_2=-1$, $\rho_1=0$.

ii) When $\rho_2\neq-1$, $\rho_1\neq0$ (or $\rho_1\neq-1$, $\rho_2\neq0$), the resulting beating soliton oscillates in each component.
The values of $\rho_1$ and $\rho_2$ affect the amplitude of soliton in each component. Performing the same process shown in Section \ref{Sec-0},
we obtain the exact solutions of beating solitons in the case of $\beta_j=\beta$. When $\rho_1=(\sqrt{3}-1)/2$, $\rho_2=(\sqrt{3}-1)/2$,
this fundamental beating-beating-beating solutions is consistent with that of the beating solitons obtained by performing the SU(3) transformation (\ref{SU(3)}) on dark-bright-bright solitons.

\section{General breather solution}\label{Sec-GB}
As shown in Section \ref{Sec-AB}, breather solutions of model (\ref{eq1}) on the plane wave (\ref{eqpw0}) under the conditions (\ref{eqwnb}) and (\ref{eqpwa}) are in fact the two-component NLSE breather solutions on the plane wave (\ref{eqsb4}).
Here, we present the explicit form of the solution describing the fundamental single breather as follows:
\begin{equation}\label{GB}
\psi_1^{(j)}=\psi_0^{(j)}\frac{G}{H},
\end{equation}
where $\psi_0^{(j)}$ denotes the plane wave (\ref{eqpw0}) under the conditions (\ref{eqwnb}) and (\ref{eqpwa}), and
\begin{eqnarray}
\begin{split}
&G=\mu_1\cos(\omega_i)+i\mu_2\sin(\omega_i)+\mu_3\cosh(\omega_i)+\mu_4\sinh(\omega_i),\nonumber\\
&H=\mu_5\cos(\omega_i)+i\mu_6\sin(\omega_i)+\mu_7\cosh(\omega_i)+\mu_8\sinh(\omega_i).
\end{split}
\end{eqnarray}
Here
\begin{equation}
\begin{split}
&\omega_r=\alpha x+\left(\chi_{ai}\gamma+\alpha(\gamma+\chi_{ar})\right)t ,\\
&\omega_i=-\gamma x+\frac{1}{2}\left(\alpha^2+2\alpha \chi_{ai}-\gamma(\gamma+2\chi_{ar})\right)t.\nonumber
\end{split}
\end{equation}
The other parameters are
\begin{eqnarray}
\begin{split}
&\mu_1=\frac{\beta_j+\chi_a^*}{(\chi_a^*-\chi_b)(\beta_j+\chi_b)}+
\frac{\beta_j+\chi_b^*}{(\chi_b^*-\chi_a)(\beta_j+\chi_a)},\\
&\mu_2=-i\frac{\beta_j+\chi_a^*}{(\chi_a^*-\chi_b)(\beta_j+\chi_b)}-
\frac{\beta_j+\chi_b^*}{(\chi_b^*-\chi_a)(\beta_j+\chi_a)},\\
&\mu_3=\frac{\beta_j+\chi_a^*}{(\chi_a^*-\chi_a)(\beta_j+\chi_a)}+
\frac{\beta_j+\chi_b^*}{(\chi_b^*-\chi_b)(\beta_j+\chi_b)},\\
&\mu_4=\frac{\beta_j+\chi_a^*}{(\chi_a^*-\chi_a)(\beta_j+\chi_a)}-
\frac{\beta_j+\chi_b^*}{(\chi_b^*-\chi_b)(\beta_j+\chi_b)},\\
&\mu_5=\frac{1}{\chi_a^*-\chi_b}+\frac{1}{\chi_b^*-\chi_a},~~
\mu_6=\frac{i}{\chi_b^*-\chi_a}-\frac{i}{\chi_a^*-\chi_b},\\
&\mu_7=\frac{1}{\chi_a^*-\chi_a}+\frac{1}{\chi_b^*-\chi_b},~~
\mu_8=\frac{1}{\chi_a^*-\chi_a}-\frac{1}{\chi_b^*-\chi_b}.\nonumber
\end{split}
\end{eqnarray}
Here, $\chi_a$ is the eigenvalue given by Eqs. (\ref{b-eqchi1234}), which is same as that of beating solitons.
In fact, the solution (\ref{GB}) follows from (\ref{eqdt1}) when we use the coefficients of the vector eigenfunctions $(c_{1,a},c_{1,b},c_{1,c},c_{1,d})=({1,1,0,0})$. Note that $\psi^{(1)}=\psi^{(3)}$, as shown in Eq. (\ref{eqsb1}).
The solutions thus in fact describe two-component general breathers ($\alpha\neq0$, $\gamma\neq0$), Akhmediev breathers ($\alpha=0$, $\gamma\neq0$), Kuznetsov-Ma breathers ($\alpha\neq0$, $\gamma=0$), and Peregrine rogue waves
($\alpha=0$, $\gamma=0$). The properties of these solutions have been studied in Refs. \cite{VKMS-f2022, VKMS-df2023, VAB-2021,VAB-Df2022,VRW-2024,VAB-2022,VRW-2022,VAB-2023,VSRB-2024}.
In this work, we focus on the coexistence of Akhmediev breathers and beating solitons with nonzero relative wavenumber.
Other cases can be easily achieved by performing the process shown here.

\section{Vector second-order beating soliton solutions}\label{Sec-2}
The exact second-order solutions of the Manakov system (\ref{eq1}) can be obtained
 in the second step of a Darboux transformations starting from the fundamental soliton solution Eq. (\ref{eqdt1}).
The corresponding eigenfunctions $\bf{\Psi}_2$=$(R_{2},S_{2},W_{2},X_{2})$ ($n=2$) of the Lax pair (\ref{UV1}) are
\begin{eqnarray}
\begin{split}
\bf{\Psi}_1=\left(
\begin{array}{cccc}
R_{1}\\
S_{1}\\
W_{1}\\
X_{1}\\
\end{array}
\right)=\mathbf{S}.\mathbf{H}.\left(
\begin{array}{cccc}
\varphi_{2,a}\\
\varphi_{2,b}\\
\varphi_{2,c}\\
\varphi_{2,d}\\
\end{array}
\right),
\end{split}
\end{eqnarray}
Then, the second-order solution can be written as
\begin{eqnarray}
&&\psi^{(1)}_2=\psi^{(1)}_1+(\lambda_2^*-\lambda_2)(\mathbf{P})_{12},\\
&&\psi^{(2)}_2=\psi^{(2)}_1+(\lambda_2^*-\lambda_2)(\mathbf{P})_{13}.\\
&&\psi^{(3)}_2=\psi^{(3)}_1+(\lambda_2^*-\lambda_2)(\mathbf{P})_{14}.
\end{eqnarray}
Here, $\psi^{(j)}_1$ denotes the vector fundamental solution, while
$(\mathbf{P})_{1i}$ represents the element of the matrix $(\mathbf{P})$ in the first row and $i$-th column, and
\begin{eqnarray}
&&\mathbf{T}=\mathbf{I}-\frac{\lambda_{1}-\lambda_{1}^*}{\lambda_2-\lambda_{1}^*}
\frac{\bf{\Psi}_{1}\bf{\Psi}_{1}^\dagger}{\bf{\Psi}_{1}^\dagger\bf{\Psi}_{1}},\\
&&\mathbf{P}=\frac{\bf{\Psi}_{2}\bf{\Psi}_{2}^\dagger}{\bf{\Psi}_{2}^\dagger\bf{\Psi}_{2}},~~~\bf{\Psi}_{2}=\mathbf{T}\bf{\Psi}_{2},
\end{eqnarray}
where $\dagger$ denotes the matrix transpose and complex conjugate.

\section{Vector soliton solutions}\label{Sec-5}

We present here the details of the construction of exact solutions of dark-drak-bright solitons and dark-bright-bright solitons via the Darboux transformation. These solutions can be used to construct beating solitons via SU(2) or SU(3) rotation symmetry.

i) We first consider the dark-dark-bright soliton solutions. We use the plane wave (\ref{eqpw0}) with the wavenumbers (\ref{eqwnb}) and the amplitudes $a_1=a_2=a$, $a_3=0$. In this case, Eq. (\ref{eqchi}) has four solutions
\begin{eqnarray}\label{5-eqchi1234}
\begin{split}		
\chi_{a,1}&=\frac{1}{2}i\left(-\alpha-i(\beta+\gamma)+\sqrt{\mu-2i\sqrt{\eta}}\right),\\
\chi_{a,2}&=\frac{1}{2}i\left(-\alpha+i(\beta+\gamma)+\sqrt{\mu-2i\sqrt{\eta}}\right),\\
\chi_{a,3}&=\frac{1}{2}i\left(-\alpha-i(\beta+\gamma)+\sqrt{\mu+2i\sqrt{\eta}}\right),\\
\chi_{a,4}&=\frac{1}{2}i\left(-\alpha+i(\beta+\gamma)+\sqrt{\mu+2i\sqrt{\eta}}\right).
\end{split}
\end{eqnarray}
where $\mu=-\beta^2+(\alpha-i\gamma)^2+4a^2\delta$, and $\eta=\beta^2(\alpha-i\gamma)^2+4a^2\beta^2\delta-4a^2\delta^2$.

For each of $\chi_{a}$, we have the corresponding spectral parameter $\lambda(\chi_{a})$ given by (\ref{eqlambda}). Substituting $\lambda(\chi_{a})$ into Eq.(\ref{det}), we obtain the fourth eigenvalue
\begin{eqnarray}\label{5-eqchi12}
\begin{split}		
\chi_{d}=-\beta.
\end{split}
\end{eqnarray}
and $\chi_c$ can be found by solving Eq.(\ref{det})  numerically.

The transformation matrix in this case is
\begin{eqnarray}\label{5-EqH3-4}
\begin{split}
&\mathbf{H}=\left(
\begin{array}{cccc}
1 & 1 & 1  & 0\\
\frac{a_1}{\beta_1+\chi_{a}} & \frac{a_1}{\beta_1+\chi_{b}} & \frac{a_1}{\beta_1+\chi_{c}} & 0\\
\frac{a_2}{\beta_2+\chi_{a}} & \frac{a_2}{\beta_2+\chi_{b}} & \frac{a_2}{\beta_2+\chi_{c}} & 0\\
0 & 0 & 0 & 1\\
\end{array}
\right).\\
\end{split}
\end{eqnarray}
The eigenfunctions $\bf{\Psi}_1$=$(R_{1},S_{1},W_{1},X_{1})$ ($n=1$) of the Lax pair (\ref{Lax-pair}) are given by
\begin{eqnarray}
\begin{split}
R_{n}&=\varphi_{n,a}+\varphi_{n,b}+\varphi_{n,c},\\
S_{n}&=\psi_{0}^{(1)}\left(\frac{\varphi_{n,a}}{\beta_1+\chi_{n,a}}+
\frac{\varphi_{n,b}}{\beta_1+\chi_{n,b}}+
\frac{\varphi_{n,c}}{\beta_1+\chi_{n,c}}\right),\\
W_{n}&=\psi_{0}^{(2)}\left(\frac{\varphi_{n,a}}{\beta_2+\chi_{n,a}}+
\frac{\varphi_{n,b}}{\beta_2+\chi_{n,b}}+
\frac{\varphi_{n,c}}{\beta_2+\chi_{n,c}}\right),\\
X_{n}&=\exp{(i\theta_3)}\left(\varphi_{n,d}\right).
\end{split}
\end{eqnarray}

ii) We then consider the dark-bright-bright soliton solutions. We still start with the plane wave (\ref{eqpw0}) with the wavenumbers (\ref{eqwnb}). However, we consider the amplitudes $a_1=a$, $a_2=a_3=0$. Then, Eq. (\ref{eqchi}) has two solutions
\begin{eqnarray}\label{5-eqchi12}
\begin{split}		
\chi_{a,1}&=\frac{1}{2}\left(-i\alpha-2\beta-\gamma-\sqrt{-(\alpha-i\gamma)^2-4a^2\delta}\right),\\
\chi_{a,2}&=\frac{1}{2}\left(-i\alpha-2\beta-\gamma+\sqrt{-(\alpha-i\gamma)^2-4a^2\delta}\right).
\end{split}
\end{eqnarray}
From (\ref{eqchi-b}), we obtain the relation between the two eigenvalues
\begin{eqnarray}\label{5-eqchi12-r}
\chi_{a,1}=-\chi_{b,2},~~~
\chi_{a,2}=-\chi_{b,1}.
\end{eqnarray}
For each of $\chi_{a}$, we have the corresponding spectral parameter $\lambda(\chi_{a})$ given by (\ref{eqlambda}). Substituting $\lambda(\chi_{a})$ into Eq.(\ref{det}), we obtain the third and fourth eigenvalue
\begin{eqnarray}\label{5-eqchiabcd}
\begin{split}		
\chi_{c}=-\beta,~~~\chi_{d}=-\beta.
\end{split}
\end{eqnarray}
We further diagonalize the matrixes $\tilde{\mathbf{U}}$ and $\tilde{\mathbf{V}}$. Namely, we have
\begin{equation}	
\varphi_{x}=\tilde{\mathbf{U}}\varphi,~~~~\varphi_{t}=\tilde{\mathbf{V}}\varphi,\label{5-uv}
\end{equation}
where the transformation matrix $\mathbf{H}$ is
\begin{eqnarray}\label{5-EqH3-3}
\begin{split}
&\mathbf{H}=\left(
\begin{array}{cccc}
1 & 1 & 0  & 0\\
\frac{a_1}{\beta_1+\chi_{a}} & \frac{a_1}{\beta_1+\chi_{b}} & 0 & 0\\
0 & 0 & 1 & 0\\
0 & 0 & 0 & 1\\
\end{array}
\right).\\
\end{split}
\end{eqnarray}
Solving Eq. (\ref{5-uv}) , we have
\begin{equation}	
\varphi_{n,l}=c_{n,l}\exp{\left[i\left(\chi_{l}x+\frac{1}{2}(\chi_{l}^2+2(1-\delta)A) t\right)\right]}.
\end{equation}
Here $c_{n,l}$ are arbitrary real constants.
Finally, the eigenfunctions $\bf{\Psi}_1$=$(R_{1},S_{1},W_{1},X_{1})$ ($n=1$) of the Lax pair (\ref{Lax-pair}) are given by
\begin{eqnarray}
\begin{split}
R_{n}&=\varphi_{n,a}+\varphi_{n,b},\\
S_{n}&=\psi_{0}^{(1)}\left(\frac{\varphi_{n,a}}{\beta_1+\chi_{n,a}}+
\frac{\varphi_{n,b}}{\beta_1+\chi_{n,b}}\right),\\
W_{n}&=\exp{(i\theta_2)}\left(\varphi_{n,c}\right),\\
X_{n}&=\exp{(i\theta_3)}\left(\varphi_{n,d}\right).
\end{split}
\end{eqnarray}
Substituting the eigenfunctions into (\ref{eqdt1}), we have the first-order soliton solutions which can be used to produce the beating solitons via
SU(2) or SU(3) rotation symmetry.

\end{appendix}

\end{document}